\documentclass[12pt]{iopart}
\usepackage{graphicx,color}
\usepackage[switch]{lineno}
\usepackage{bm}
\usepackage{bbm}
\usepackage{mathrsfs}
\usepackage{cite}
\usepackage{url}
\usepackage{iopams}
\usepackage{bbold}
\usepackage{soul} 
\usepackage{perpage}
\setstcolor{blue}
\def\Tr{\mbox{Tr}\,}

\newcommand{\rmnum}[1]{\romannumeral #1}
\newcommand{\hX}{\hat{X}}

\newcommand{\parlr}[1]{\left( #1 \right )}
\newcommand{\sqlr}[1]{\left[ #1 \right]}
\newcommand{\flrlr}[1]{\left\{ #1 \right\}}
\newcommand{\abs}[1]{\left| #1 \right |}

\makeatletter
\newcommand{\Rmnum}[1]{\expandafter\@slowromancap\romannumeral #1@}
\makeatother
   \newcommand{\ket}[1]{\left|#1\right>} \newcommand{\bra}[1]{\left<#1\right|}
   
   \DeclareMathAlphabet{\mathcalligra}{T1}{calligra}{m}{n}
   \newcommand{\bx}{\mathbf{x}}
   \newcommand{\hsig}{\sigma}

\begin{document}

\title{Quantum fluctuation theorems and power measurements}

\author{B. Prasanna Venkatesh$^1$ \footnote{Present address: Institute for Quantum Optics and Quantum Information of the Austrian Academy of Sciences, Technikerstra\ss e 21a, Innsbruck 6020, Austria},  Gentaro Watanabe$^{1,2,5}$ and Peter Talkner$^{1,3,4}$}
\address{$^1$Asia Pacific Center for Theoretical Physics (APCTP), San 31, Hyoja-dong, Nam-gu, Pohang, Gyeongbuk 790-784, Korea}
\address{$^2$Department of Physics, POSTECH, San 31, Hyoja-dong, Nam-gu, Pohang,
Gyeongbuk 790-784, Korea}
\address{$^3$Institut f\"{u}r Physik, Universit\"{a}t Augsburg, Universit\"{a}tsstra\ss e 1, D-86135 Augsburg, Germany}
\address{$^4$Institute of Physics, University of Silesia, 40007 Katowice, Poland}
\address{$^5$Center for Theoretical Physics of Complex Systems, Institute for Basic Science (IBS), Yuseong-gu, Daejeon 305-811 Korea}
\ead{balasubv@apctp.org}

\date{\today}
\begin{abstract}
Work in the paradigm of quantum fluctuation theorems of Crooks and Jarzynski, is determined by projective measurements of energy at the beginning and end of the force protocol. In analogy to classical systems, we consider an alternate definition of work given by the integral of the supplied power determined by integrating up the results of repeated measurements of the instantaneous power during the force protocol. We observe that such a definition of work, in spite of taking account of the process dependence, has different possible values and statistics from the work determined by the conventional two energy measurement approach (TEMA). In the limit of many projective measurements of power, the system's dynamics is frozen in the power measurement basis due to the quantum Zeno effect leading to statistics only trivially dependent on the force protocol. In general the Jarzynski relation is not satisfied except for the case when the instantaneous power operator commutes with the total Hamiltonian at all times. We also consider properties of the joint statistics of power-based definition of work and TEMA work in protocols where both values are determined. This allows us to quantify their correlations. Relaxing the projective measurement condition, weak continuous measurements of power are considered within the stochastic master equation formalism. Even in this scenario the power-based work statistics is in general not able to reproduce qualitative features of the TEMA work statistics.
\end{abstract}
\maketitle
\section{Introduction}\label{sec:intro}
Transient fluctuation theorems are exact relations restricting the statistics of work performed by externally controlled classical forces. While the considered system initially must be in a thermal equilibrium state, the subsequent forcing may drive it to out of equilibrium to states that cannot be described in terms of linear response theory. Exact relations of such form were pioneered by Bochkov and Kuzovlev \cite{Bochkov77} and the ones pertinent to our discussion bear the names of Jarzynski \cite{Jarzynski97} and Crooks \cite{Crooks99}. 
The Crooks relation \cite{Crooks99},
\begin{equation}
p_{\Lambda}(w) = e^{-\beta(\Delta F - w)} p_{\bar{\Lambda}}(-w) \label{eq:crooks}
\end{equation}
relates the probability density function (pdf), $p_{\Lambda}(w)$, of work performed on a system with Hamiltonian $H[\lambda(t)]$ by the action of a generalized force $\lambda(t)$ that varies according to the protocol  $\Lambda = \{ \lambda(t)|0\leq t \leq \tau \}$ to the work pdf of another process governed by the time-reversed protocol $\bar{\Lambda} = \{\lambda(\tau-t)|0 \leq t \leq \tau\}$\footnote{For the sake of simplicity we restrict ourselves to forces $\lambda(t)$ which transform evenly under time-reversal.}. Both processes start in a canonical equilibrium state at the same inverse temperature $\beta$ described by the density matrices $\rho_t = Z^{-1}(t) e^{-\beta H[\lambda(t)]}$ with $t=0$ and $\tau$ for the forward and the backward process, respectively. Here $Z(t) = \Tr e^{-\beta H[\lambda(t)]}$, for $t=0$ and $\tau$ yields the partition functions which determine the free energy difference $\Delta F = -\beta^{-1} \ln (Z(\tau)/Z(0))$. 

A straightforward integration of (\ref{eq:crooks}) with $e^{-\beta w}$ brought to its left hand side 
leads to the Jarzynski equality \cite{Jarzynski97},
\begin{eqnarray}
\langle e^{-\beta w} \rangle = e^{-\beta \Delta F} \label{eq:jarzynski}.
\end{eqnarray}
Fluctuation theorems have been shown to be valid in a variety of situations ranging from open classical systems \cite{Jarzynski04}, closed \cite{Kurchan00,Tasaki00,Talkner07} and open quantum systems \cite{Crooks08,Talkner09,Campisi09}. For recent reviews of this topic  \cite{Jarzynskirev11,Seifert12} (focus on classical systems) and \cite{Esposito09,Campisirev11,Campisireverr11,Hanggi15} (focus on quantum systems) can be consulted.

These theorems have also been verified experimentally for a variety of classical systems \cite{Hummer01,Liphardt01,Trepagnier04,Ciliberto05} and have also been used to determine free energy changes in molecular systems by measuring work statistics in controlled non-equilibrium processes \cite{Hummer01,Liphardt01}. Further details on this topic are available in the reviews \cite{Bustamante05,Ritort08}. 

The situation regarding experimental verification of fluctuation theorems for quantum systems is still in its nascent stages. One central issue in the quantum context is that work is not an observable \cite{Talkner07}; it is process dependent and therefore the potential values that it may take in a single experiment cannot be represented as the eigenvalues of a hermitian operator acting on the Hilbert space of the considered system. This issue underpins the most common approach to determine work statistics for a quantum system, the so-called two-energy-measurement approach (TEMA) \cite{Talkner07,Campisirev11,Campisireverr11}: here the work supplied by a time dependent force $\lambda(t)$ during the time interval $0\leq t \leq \tau$ is defined as the difference between the system's energy at the final and initial times. The TEMA can be divided into the following steps: (\rmnum{1}) the system is prepared in a thermal equilibrium state at inverse temperature $\beta$ at $t=0$ with the initial Hamiltonian $H[\lambda(0)]$; (\rmnum{2}) a projective measurement of energy is performed yielding one of the eigenvalues $e_n(0)$ of $H[\lambda(0)]$ as possible outcome; due to the measurement the state of the system is projected to the associated energy eigenstate $\vert n; 0 \rangle$; (\rmnum{3}) subsequently it undergoes a unitary time evolution during $0 \leq t \leq \tau$ governed by the Hamiltonian $H[\lambda(t)]$ which changes in time according to the prescribed protocol $\Lambda$; (\rmnum{4}) at $t = \tau$ a second projective measurement of energy is performed; its outcome is an eigenvalue $e_m(\tau)$ of the final Hamiltonian $H[\lambda(\tau)]$. The work done in such a process is given by $w_e = e_m(\tau) - e_n(0)$. In general, projective measurements turn out to be difficult to perform in an experiment. Nonetheless proposals for implementing  such a scheme using trapped ions was suggested in \cite{Huber08} and its experimental realisation was recently reported in \cite{Kim14}. In an indirect verification of the fluctuation theorems detailed in \cite{Dorner13,Mazzola13,Campisi13}, the experimental difficulties of projective measurements were circumvented by encoding the characteristic function of work in the reduced state of an ancillary system that interacts with the system of interest with a strength determined by the force protocol. An experiment, using NMR spectroscopy, of such a proposal was reported in \cite{Batalhao13}. A second way to address the difficulty of performing projective measurements is to ask if the replacement of the same by non-projective generalized measurements, which may be easier to implement in an experiment, preserves the fluctuation theorems \cite{Venkatesh14,Watanabe14} (see also \cite{Roncaglia14, deChiara14} for a recent proposal to measure work as the outcome of a single generalised measurement). With this line of enquiry we found in an earlier publication \cite{Venkatesh14} a no-go theorem restricting projective measurements of energy as the only ones within a large class of generalized measurements that satisfy both Crooks and Jarzynski equalities for arbitrary protocols. Although, for some specific forms of the generalized measurements, modified fluctuation theorems may be derived \cite{Watanabe14}. 

In this work, we focus on yet another way to address the issue of measurement of work in quantum systems. Conventionally, in classical systems work can be determined in an incremental way by integrating up the supplied power which can be inferred from the instantaneous state of the system \cite{Bustamante05}. A direct extension of such a method to quantum systems is difficult since the system will have to be continuously monitored \cite{Misra77,Itano90}. It is known that a continuous monitoring of a quantum system can cause the freezing of coherent dynamics, a phenomenon known as the quantum Zeno effect \cite{Misra77,Itano90}. The central aim of this work is to explore in detail how and why the extension of power measurements to estimate the work done on quantum systems fails to provide 
work statistics that satisfies the transient fluctuation theorems of Crooks \cite{Crooks99} and Jarzynski \cite{Jarzynski97}.

The paper is organized as follows. In section \ref{sec:powmeas} we define the quantum mechanical variable corresponding to the instantaneous power supplied to a system that experiences a force via a coupling to a generalized coordinate. We introduce an estimate of work computed by incremental projective measurements of the power in conjunction with the protocol, and explain how such an estimate differs from the usual TEMA based work both in magnitude and range of the possible values. In the limit of frequent measurements, the system's unitary evolution is frozen due to the quantum Zeno effect (we refer to such a regime of the dynamics as the ``Zeno limit" henceforth). We analyze the statistics of the power-based work in this limit and point out how the commutator of the power operator with the total Hamiltonian is crucial in deciding the merits of the work estimate. We also derive an inequality for the dissipated work estimated by the power measurement in the Zeno limit. Further, the results so far obtained are illustrated by the example of a two-level system undergoing an avoided crossing. In section \ref{sec:jointstat}, we allow for energy measurements at the beginning and the end of the force protocol additionally to the coordinate measurements necessary for the power measurements and compare the distributions of TEMA, $w_e$, and power-based work  estimates $w_p$. In section \ref{sec:fintime} we relax the assumption of projective instantaneous measurements of the power operator and adopt a weak, continuous measurement of power to define the work. The treatment there is carried out within the framework of Stochastic Master Equations (SME) \cite{Jacobs06,WM}. Work statistics obtained from numerical solution of SMEs for the two-level system example are also studied. In section \ref{sec:conclusion} we conclude the paper. Appendices \ref{app:genpower} and \ref{app:smederiv} provide additional details omitted from the main text of the paper.

\section{Power measurement and the quantum Zeno effect} \label{sec:powmeas}
In classical mechanics, the energy-work relation can be invoked to define the work supplied to a thermally closed system\footnote{We denote a system as thermally closed if it does not exchange energy and/or particles with its environment. In order to be able to perform work on it,
it must though be possible to externally change some of the system parameters.} as the increase in its energy. First, let us consider a classical system whose energy is determined by the Hamiltonian $H^{\mathrm{cl}}[\mathbf{z},\lambda(t)]$  where $\mathbf{z}$ is a point in phase-space and $\lambda(t)$ is an external parameter which is varied in time leading to a change in the system's energy. Hence the work done by changing $\lambda(t)$ in the interval $0\leq t \leq \tau$ according to some specific protocol is given by \cite{Jarzynski97}
\begin{eqnarray}
w^{\mathrm{cl}} = H^{\mathrm{cl}} \sqlr{\mathbf{Z(\tau,\mathbf{z})},\lambda(\tau)} - H^{\mathrm{cl}}\sqlr{\mathbf{z},\lambda(0)} \label{eq:ediffwclass},
\end{eqnarray}
where $\mathbf{Z}(t,\mathbf{z})$ is the solution of Hamilton's equation of motion
\begin{eqnarray*}
\frac{d}{dt} \mathbf{Z}  = \flrlr{H^{\mathrm{cl}}\sqlr{\mathbf{Z},\lambda(t)},\mathbf{Z}}
\end{eqnarray*}
 at time $t$ for a trajectory with the initial condition $\mathbf{Z}(0,\mathbf{z}) = \mathbf{z}$. (\ref{eq:ediffwclass}) can be rewritten as an integral of the total time-derivative of the Hamiltonian which agrees with the partial time-derivative. This gives an equivalent expression for work,
 \begin{eqnarray}
 w^{\mathrm{cl}} = \int_0^{\tau} dt \frac{\partial H^{\mathrm{cl}}\sqlr{\mathbf{Z}(t,\mathbf{z}),\lambda(t)}}{\partial t} \label{eq:powintwclass},
 \end{eqnarray}
 where the integrand is the instantaneous power $L^{\mathrm{cl}}$ supplied to the system at time $t$, i.e.
 \begin{eqnarray}
 L^{\mathrm{cl}}(t,\mathbf{z}) &= \frac{\partial H\sqlr{\mathbf{Z}(t,\mathbf{z}),\lambda(t)}}{\partial t}\\
&=\frac{\partial H\sqlr{\mathbf{Z}(t,\mathbf{z}),\lambda(t)}}{\partial \lambda} \frac{d \lambda(t)}{d t}
 \label{eq:classpow}.
 \end{eqnarray}
 The power-based work expression (\ref{eq:powintwclass}) is more convenient to determine work in experiments \cite{Hummer01}. Since the initial condition $\mathbf{z}$ is only specified in terms of a probability distribution (for, e.g., a canonical distribution if the system is initially in thermal equilibrium), work becomes a random quantity.

 In quantum mechanics, both expressions (\ref{eq:ediffwclass}) and (\ref{eq:powintwclass}) can in principle be extended to operational definitions of work which, however, turn out to be no longer equivalent to each other. Here, we note that there have been earlier attempts \cite{Chernyak04,Engel07,Liu12,Wang13,Solinas13,Suomela14} to define work in quantum systems using expressions analogous to (\ref{eq:powintwclass}). In \cite{Chernyak04} this is done in the context of a driven harmonic oscillator interrupted by a small number of measurements of the coordinate and in \cite{Wang13} the same system is examined under the continuous quantum histories framework for the power operator. In the context of superconducting Cooper-pair box systems some lower order moments of a power-based work were considered in \cite{Solinas13} for a closed system and extended to include open systems in \cite{Suomela14}. In some of these attempts \cite{Engel07,Solinas13}, power-based work has been treated as a standard quantum mechanical observable (with a corresponding operator) in contrast to the view point in \cite{Talkner07}, which we adopt. One central aim of this work is to show that even a more careful implementation of the power-based work for quantum systems taking into account the process dependence will generally lead to qualitatively different statistics from the TEMA definition of work and generally fail to satisfy the fluctuation relations of Crooks and Jarzynski.
 
 To that end, we consider the simplest possible situation where a single scalar parameter $\lambda(t)$ acts as a force on a system via a generalized coordinate $X$.  In the absence of this force the system is described by the Hamiltonian $H_0$. In the quantum scenario, the generalised coordinate $X$ corresponds to a self-adjoint operator $\hX$ acting on the system's Hilbert space. The total Hamiltonian is then given by:
\begin{eqnarray}
H[\lambda(t)] = H_0 + \lambda(t) \hX. \label{eq:totham}
\end{eqnarray}
In analogy to the classical form of power (\ref{eq:classpow}) we define the power operator $L(t)$ as
\begin{eqnarray}
L(t) \equiv \frac{\partial H^{H}[\lambda(t)]}{\partial t} =  \dot{\lambda}(t) \hX(t) \label{eq:powerop},
\end{eqnarray}
where the superscript $H$ denotes the Heisenberg picture and the dot a time derivative. Since the generalised coordinate does not explicitly depend on time, we simply indicate the Heisenberg picture by the presence of the time-argument omitting the superscript $H$. In order to determine the power, one needs to perform a measurement of the generalized coordinate $\hX$. The possible measurement outcomes, in a projective measurement \cite{vonneumann96}, are determined by the eigenvalues $x_{\alpha}$ of the generalized coordinate $\hX =  \sum_{\alpha} x_{\alpha} \Pi_{\alpha}^X$. For the sake of simplicity we assume that $\hX$ has a non-degenerate discrete spectrum. Hence, the eigenprojection operators $\Pi_{\alpha}^X = \ket{\varphi_{\alpha}} \bra{\varphi_\alpha}$ are determined by the eigenfunction $\ket{\varphi_{\alpha}}$ of $\hX$. In order to precisely capture the work as the integral of the power one should, in principle, continuously measure the generalized coordinate. This though will inevitably freeze the dynamics of the system in an eigenstate of $\hX$. Provided that the force protocol is sufficiently slow compared to the unitary dynamics of the system, one can try to avoid the full halt of the dynamics by performing only  a finite number of power measurements and approximate the integral by a discrete sum:  
\begin{eqnarray}
w^{(N)}_p = \displaystyle \sum_{i=1}^N \dot{\lambda}(t_i) x_{\alpha_i} h, \label{eq:workpowbN}
\end{eqnarray}
where we assumed that $N$ measurements take place at regularly spaced times $t_i$ with $t_{i+1}-t_i = t_1 = h$ in the interval $\parlr{t_0 \equiv 0}\leq t \leq \parlr{t_{N+1} \equiv \tau}$. The first basic difference between the power-based and TEMA work estimates emerges from the varying allowed values of work that the two approaches produce. On the one hand, the set of possible TEMA work values is given by 
\begin{eqnarray}
\mathcal{W}^{e} = \flrlr{w = e_m(\tau)-e_n(0)|m\in I(\tau),n\in I(0)}, \label{eq:workTMA}
\end{eqnarray}
where $I(t)$ is the set of indices labelling the spectrum of $H(t)=\sum_{n \in I(t)} e_n(t) \Pi_n(t)$. Here $\Pi_n(t)$ are operators projecting on the eigenstates corresponding to $e_n(t)$. Comparing with all possible power-based work values having the form given by (\ref{eq:workpowbN}), it is apparent that both the number of possible values and magnitudes of work are different in the two approaches. Indeed for a system with finite $D$-dimensional Hilbert space, the maximum number of possible work values from TEMA is $D^2$, whereas in the power measurement case with $N$ measurements it is given by $D^{N}$. Whereas the allowed TEMA work values are functions of all parameters entering the full Hamiltonian (\ref{eq:totham}), the power-based work values depend solely on the time-derivatives of the force at the measurement times. Only the probabilities which characterize the occurrence of these work values may depend on the other parameters of the system.

Still one could hope that the two estimates of work might be similar in a statistical sense. For this purpose we now compute the distribution for the work estimate in (\ref{eq:workpowbN}) to compare it  with the distribution of the TEMA work values in terms of their cumulative probabilities. Beginning at $t=0$ with a canonical density matrix $\rho_0 = Z^{-1}(0) e^{-\beta H[\lambda(0)]}$ we obtain for the joint probability $\mathscr{P}_{\Lambda}(\mathbf{x})$ of finding the eigenvalues $\mathbf{x} = \parlr{x_{\alpha_1}, x_{\alpha_2}, \cdots, x_{\alpha_N}}$ in the $N$ measurements at times $t_1,t_2, \cdots, t_N$:
\begin{eqnarray}
\mathscr{P}_{\Lambda}(\mathbf{x}) = \Tr V_{\Lambda}\parlr{\bx} \rho_0 V_{\Lambda}^{\dagger}\parlr{\bx} \label{eq:jointprobproj},
\end{eqnarray}
where
\begin{eqnarray}
V_{\Lambda}\parlr{\bx} = \Pi_{\alpha_{N}}^X U_N \Pi_{\alpha_{N-1}}^X \cdots U_2 \Pi_{\alpha_{1}}^X U_1 \label{eq:jointprobform}.
\end{eqnarray}
Here $U_k = U_{t_k,t_{k-1}}(\Lambda)$ denotes the time-evolution operator for the system from time $t_{k-1}$ to $t_k$ under the influence of the protocol $\Lambda$. It is the solution to the Schr\"{o}dinger equation
\begin{eqnarray}
i\hbar \frac{\partial U_{t,s}}{\partial t} = H[\lambda(t)] U_{t,s}(\Lambda), 
\label{U}
\end{eqnarray}
obeying the initial condition
\begin{eqnarray*}
U_{s,s}(\Lambda) = \mathbb{1}.
\end{eqnarray*}

As evident from (\ref{eq:jointprobform}), we have taken the measurements of the generalized coordinate to be projective (following von Neumann \cite{vonneumann96}). We will relax this assumption in Appendix \ref{app:genpower} and consider generalised measurements of the coordinate. Expressing the projection operators in terms of the eigenfunctions of $\hX$, the joint probability becomes:
\begin{eqnarray}
\mathscr{P}_{\Lambda}(\mathbf{x}) = \displaystyle \prod_{k=1}^{N-1} \left | \bra{\varphi_{\alpha_{k+1}}} U_{k+1} \ket{\varphi_{\alpha_{k}}}\right |^2 \bra{\varphi_{\alpha_1}} U_1 \rho_0 U^{\dagger}_1 \ket{\varphi_{\alpha_1}} \label{eq:jointprobeig}.
\end{eqnarray}
The pdf $p_{\Lambda p}^{(N)} (w)$ of finding the value $w$ for the power-based work estimate $w_p^{(N)}$ is then given by:
\begin{eqnarray}
p_{\Lambda p}^{(N)} (w) =\displaystyle \sum_{\flrlr{{\alpha_i}}} \delta \parlr{w-\sum_{i=1}^{N} \dot{\lambda}(t_i)x_{\alpha_i}h} \mathscr{P}_{\Lambda}(\bx). \label{eq:projfinmeaspowpdf}
\end{eqnarray}
For a large number of measurements, the Zeno effect \cite{Misra77,Itano90} causes a freezing of the system in the state corresponding to the outcome of the first measured eigenvalue of $\hX$. This is a consequence of the fact that the transition probabilities $\left | \bra{\varphi_{\alpha_{k+1}}} U_{k+1} \ket{\varphi_{\alpha_{k}}}\right |^2$ between different eigenstates of $\hX$ vanish as $h^2$. Hence for large values of $N$ (which correspond to small values of $h = \tau/(N+1)$), the joint probability in (\ref{eq:jointprobeig}) becomes
\begin{eqnarray}
\mathscr{P}_{\Lambda}(\mathbf{x}) = \displaystyle \prod_{k=1}^{N-1} \delta_{x_{\alpha_k},x_{\alpha_{k+1}}} \bra{\varphi_{\alpha_1}} \rho_0 \ket{\varphi_{\alpha_1}} + O(h).
\end{eqnarray}
Putting this asymptotic result into the expression in (\ref{eq:projfinmeaspowpdf}) we obtain 
\begin{eqnarray}
p_{\Lambda p}^{(\infty)}(w) = \displaystyle \sum_{\alpha} \delta \parlr{w - \sqlr{\lambda(\tau)-\lambda(0)}x_{\alpha}}\bra{\varphi_{\alpha}}\rho_0 \ket{\varphi_{\alpha}}, \label{eq:projinfmeaspowpdf}
\end{eqnarray}
for the work pdf neglecting corrections of the order $h$. Note that in the above $h \rightarrow 0$ limit, the sum in the delta function specifying the work in (\ref{eq:projfinmeaspowpdf}) can be replaced by an integral yielding $\lim_{N \to \infty}\sum_{i=1}^{N} \dot{\lambda}(t_i)x_{\alpha_1}h = \sqlr{\lambda(\tau)-\lambda(0)}x_{\alpha_1}$.

Having obtained the pdf for the power-based estimate of work, we can immediately check if the Jarzynski equality is satisfied. To that end we have for the average of the exponentiated work the following expression:
\begin{eqnarray}
\langle e^{-\beta w} \rangle_p = \int dw\, p_{\Lambda p}^{\infty}(w) e^{-\beta w} = \frac{\Tr e^{-\beta \sqlr{\lambda(\tau) -\lambda(0)} \hX} e^{-\beta\sqlr{H_0 + \lambda(0) \hX}}}{Z(0)} \label{eq:expbetawpowinf}.
\end{eqnarray}
For the Jarzynski equality to hold, the numerator of the right-hand side must coincide with the partition function $\Tr e^{-\beta H[\lambda(\tau)]}$. However, this is only the case if the unperturbed Hamiltonian $H_0$ commutes with the generalized coordinate $\hX$, i.e. if  $[ H_0, \hX ] = 0$, which is an atypical situation. Because in the commuting case, the forcing does not lead to transitions between different eigenstates of $H[\lambda(t)]$, the set of allowed TEMA work values is given by
\begin{equation}
\mathcal{W}^e = \{ (\lambda(\tau)  - \lambda(0)) x_\alpha|\alpha \in I \}
\label{Wec}
\end{equation}
and hence coincides with the set of allowed power-based work values. This follows from the eigenvalues of $H\sqlr{\lambda(t)}$ taking the form $e_{\alpha}(t) = e_{\alpha} + \lambda(t) x_{\alpha}$, $\alpha \in I$ where
$I$ is the time-independent set labelling the eigenvalues $e_{\alpha}$ and $x_{\alpha}$ of $H_0$ and $\hX$, respectively.

In general, when $H_0$ and $\hX$ do not commute, the power-based work estimate does not conform with the Jarzynski equality\footnote{In \cite{Wang13} a similar conclusion was obtained using the continuous quantum histories framework for the model system of a center-shifted harmonic oscillator.}. In this case one finds with the Golden-Thompson inequality \cite{Golden65} $\Tr e^{A}e^B \geq \Tr e^{A+B}$ (with $A$ and $B$ being hermitian) in (\ref{eq:expbetawpowinf}) yielding
\begin{eqnarray*}
\langle e^{-\beta w} \rangle_p \geq \frac{\Tr e^{-\beta \sqlr{\lambda(\tau) -\lambda(0)} \hX-\beta\sqlr{H_0 + \lambda(0) \hX}}}{Z(0)} = e^{-\beta \Delta F}.
\end{eqnarray*}
Thus the power-based work estimate satisfies the inequality
\begin{eqnarray}
\langle e^{-\beta \parlr{w-\Delta F}} \rangle_p \geq 1.
\end{eqnarray}
Hence, an estimate of the free energy based on the exponentiated power-based work average underestimates the true value
\begin{eqnarray}
\Delta F_p \equiv -\beta^{-1} \ln \langle e^{-\beta w} \rangle_p \leq \Delta F.
\end{eqnarray}
A more detailed comparison can be made on the basis of the power-based work pdf (\ref{eq:projinfmeaspowpdf}) and the corresponding TEMA work pdf $p_{\Lambda e}(w)$ given by
\cite{Campisirev11,Campisireverr11}:
\begin{eqnarray}
p_{\Lambda e}(w) = \sum_{m,n} \delta \parlr{w-e_m(\tau) + e_n(0)} p_{\Lambda}(m,n),
\label{pe}
\end{eqnarray}
where the joint probability $p_{\Lambda}(m,n)$ to find the eigenstates $\ket{n;0}$ and $\ket{m;\tau}$ in projective energy measurements at the beginning and the end of the force protocol, respectively, reads
\begin{eqnarray}
p_{\Lambda}(m,n) = \Tr \Pi_m (\tau) U_{\tau,0}(\Lambda) \Pi_n(0) \rho_0 \Pi_n(0)U_{\tau,0}^{\dagger}(\Lambda).
\end{eqnarray} 
In the above equation $\Pi_j(t)$ denotes the eigenprojection operators corresponding to the eigenvalue $e_j(t)$ of the Hamiltonian $H\sqlr{\lambda(t)}$ and $U_{\tau,0}(\Lambda)$ is the unitary time evolution operator from the beginning to the end of the force protocol defined in (\ref{U}). 

We further elucidate the differences of the work pdfs, (\ref{eq:projfinmeaspowpdf}) and (\ref{pe}), which both are rather involved, by means of a simple example.

\subsection{Landau-Zener}
Next we want to illustrate the differences between the power-based estimates and the TEMA work values
and also the approach to the Zeno limit for a driven two-level system undergoing an avoided level-crossing.

The Hamiltonian for a two-level system driven through an avoided crossing, known as the Landau-Zener(-St\"{u}ckelberg-Majorana) model \cite{LZSM}, is given by
\begin{eqnarray}
H_{LZ}(t) = \Delta \sigma_x + \frac{v t}{2}\sigma_z \label{eq:LZmodel},
\end{eqnarray}
where $\sigma_j$ are the Pauli matrices. In this case the power operator is given by $L = v \sigma_z/2$. Thus the power measurement involves projective measurements in the $\sigma_z$ basis. Digressing from our previous convention, we follow the usual custom and consider a symmetric protocol about $t=0$ that runs between $-\tau/2 \leq t \leq \tau/2$. The Hamiltonian (\ref{eq:LZmodel}) is readily diagonalized yielding for the eigenvalues $e_j(t) = (-1)^j \sqrt{(vt/2)^2 + \Delta^2}$ with $j=1,2$. Hence, the possible TEMA work values are given by:
\begin{eqnarray}
\mathcal{W}^{e} = \flrlr{-E_{\mathrm{max}},0,E_{\mathrm{max}}}, \label{eq:weLZ}
\end{eqnarray}
with $E_{\mathrm{max}} = 2 \sqrt{  (v\tau/4)^2 + \Delta^2   }$, whereas the set of possible work values based on $N$ power measurements becomes 
\begin{eqnarray}
\mathcal{W}^p = \flrlr{\frac{v \tau}{2(N+1)} g,\,\,\,\, g =-N,-N+2,\cdots , N},  \label{eq:wpLZ}
\end{eqnarray}
clearly showing a fundamental difference between the two approaches. We also note here that for this example the range of TEMA work bounds the possible values in the power measurement estimate. The range of the latter increases with the number $N$ reaching $\pm v \tau/2$ for $N\to \infty$. The magnitude of the maximum work value possible from the TEMA, $E_{\mathrm{max}}$, is always larger than $v \tau/2$ but in the limit of large enough $\tau$ such that $v \tau/4 \gg \Delta$, it approaches the latter.

According to (\ref{eq:projinfmeaspowpdf}), the work pdf for the LZ problem estimated by power measurements in the Zeno limit becomes
\begin{eqnarray}
p_{\Lambda p}^{(\infty)}(w) = \bra{z_+} \rho_0 \ket{z_+} \delta(w-v\tau/2) + \bra{z_-} \rho_0 \ket{z_-} \delta(w+v\tau/2) \label{eq:LZinfpowpdf},
\end{eqnarray}
where $\ket{z_{\pm}}$ denote the eigenstates of $\sigma_z$ and $\rho_0 = e^{-\beta H_{LZ}(-\tau/2)}/Z(-\tau/2)$ is the initial density matrix. On the other hand, the work pdf from the TEMA is given by \cite{Pekola13}
\begin{eqnarray}
p_{\Lambda e}(w) = p_e P_{LZ}\delta(w+E_{\mathrm{max}}) + (1-P_{LZ})\delta(w) + p_g P_{LZ} \delta(w-E_{\mathrm{max}}),
\end{eqnarray}
with $p_g = \parlr{1+e^{-\beta E_{\mathrm{max}}}}^{-1}$, $p_e = 1-p_g$ and $P_{LZ}= e^{-2 \pi \Delta^2/(\hbar v)}$ \footnote{Strictly speaking this expression is valid only for $\tau \rightarrow \infty$ but provides a very good approximation for large finite $\tau$ with $v \tau/4 \gg \Delta $.} denotes the LZ tunneling probability \cite{LZSM} from the ground state at $t=-\tau/2$ to the excited state at $t=\tau/2$.
\begin{figure} 
\centering
\includegraphics[width=0.6\textwidth]{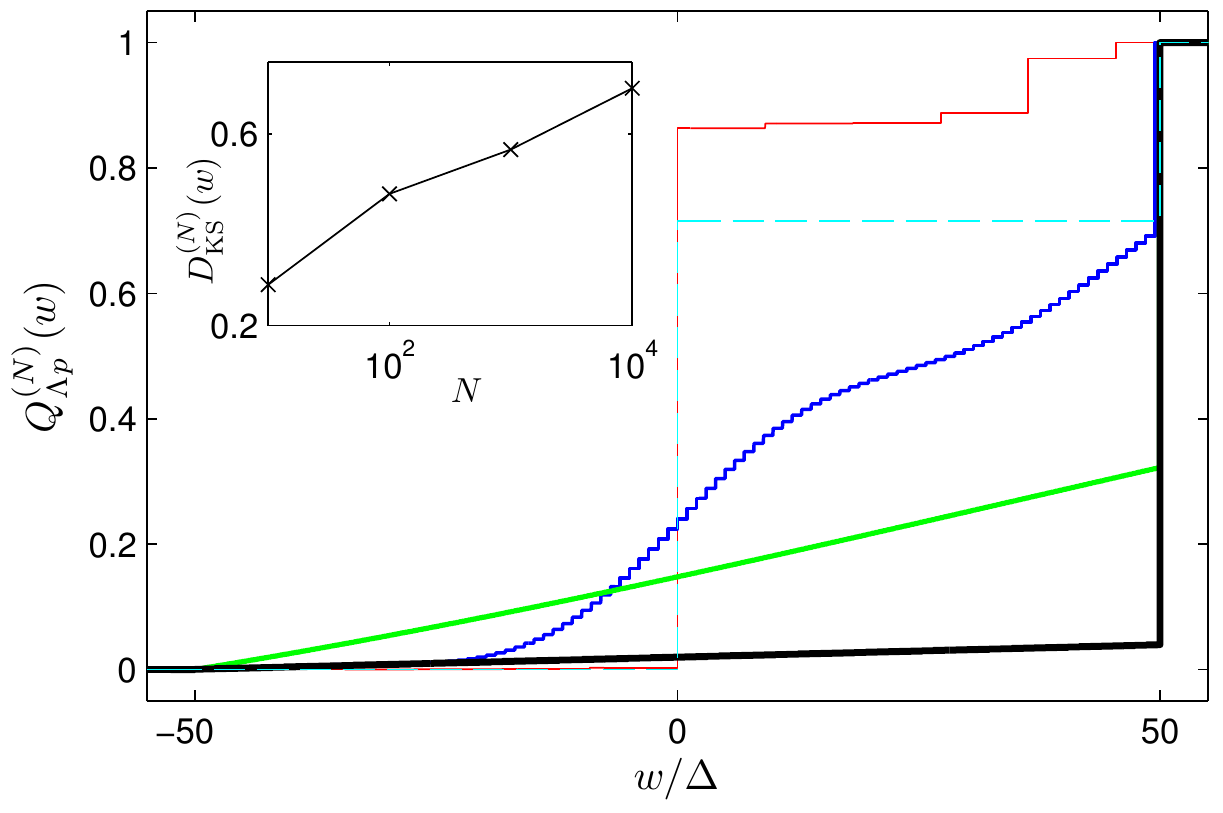}
\caption{Cumulative probability of work estimated from the sum of $N$ power measurements for the LZ problem with $v = 5 \Delta^2/\hbar$ and $\tau = 20 \hbar/\Delta$ ($N = 10$ - red, $N=10^2$ - blue, $N=10^3$ - green, and $N=10^4$ - black with ascending line thickness). The initial temperature is small $\beta E_{\mathrm{max}} = 10$. For large $N$, due to the Zeno effect, a distinct peak appears at the maximum value of $w$ (see text for discussion). For comparison the pdf computed from TEMA is also shown (cyan dashed line). The inset shows the Kolmogorov-Smirnov distance between the TEMA and power-based work estimate as a function of $N$.}
\label{fig:par1betahighzenoapp}
\end{figure}

\begin{figure}
\centering 
\includegraphics[width=0.6\textwidth]{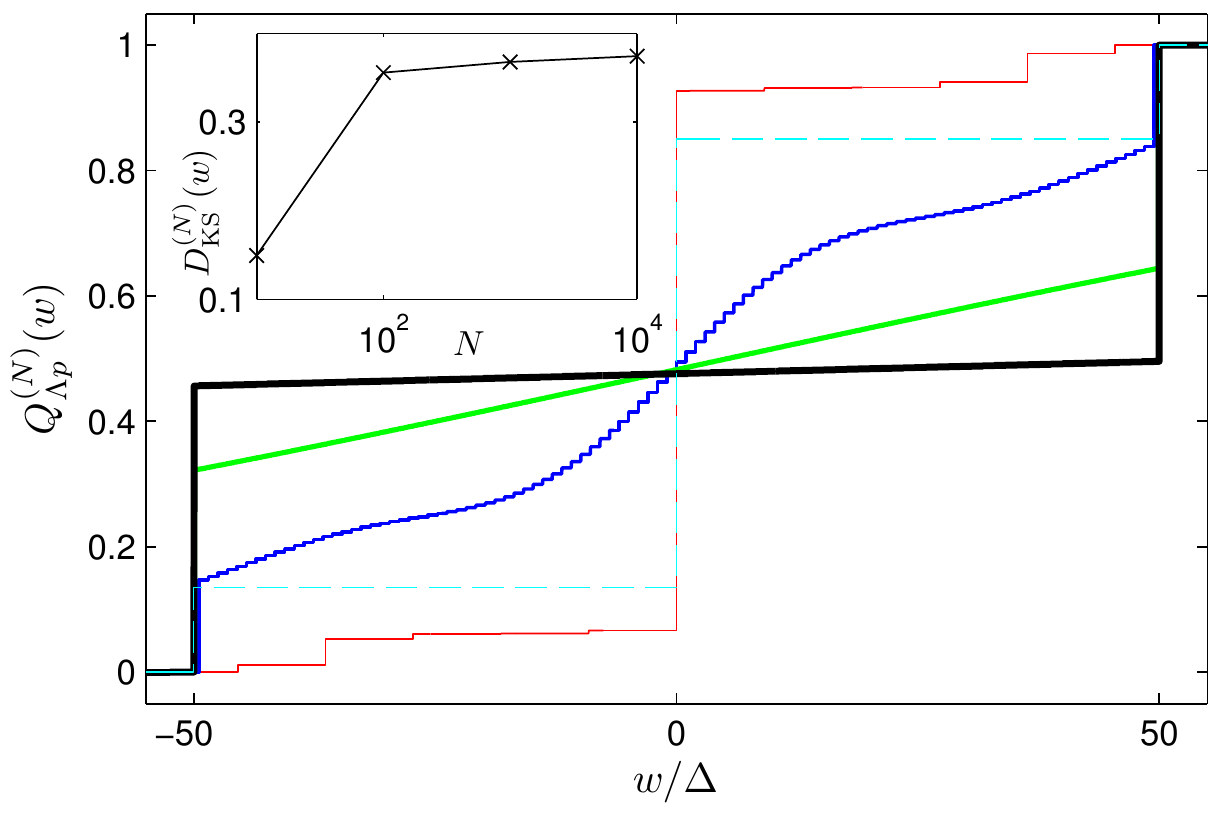}
\caption{Cumulative probability of work estimated from the sum of $N$ power measurements for the LZ problem for the same system sweep parameters (and same legends) as in figure \ref{fig:par1betahighzenoapp}. The initial temperature is large $\beta E_{\mathrm{max}} = 10^{-1}$. For large $N$, due to the Zeno effect, two distinct peaks appear at the maximum and minimum value of $w$ (see text for discussion). For comparison the cumulative probability computed from TEMA is also shown (cyan dashed line). The inset shows the Kolmogorov-Smirnov distance between the TEMA and power-based work estimate as a function of $N$. }
\label{fig:par1betalowzenoapp}
\end{figure}

\begin{figure}
\centering 
\includegraphics[width=0.6\textwidth]{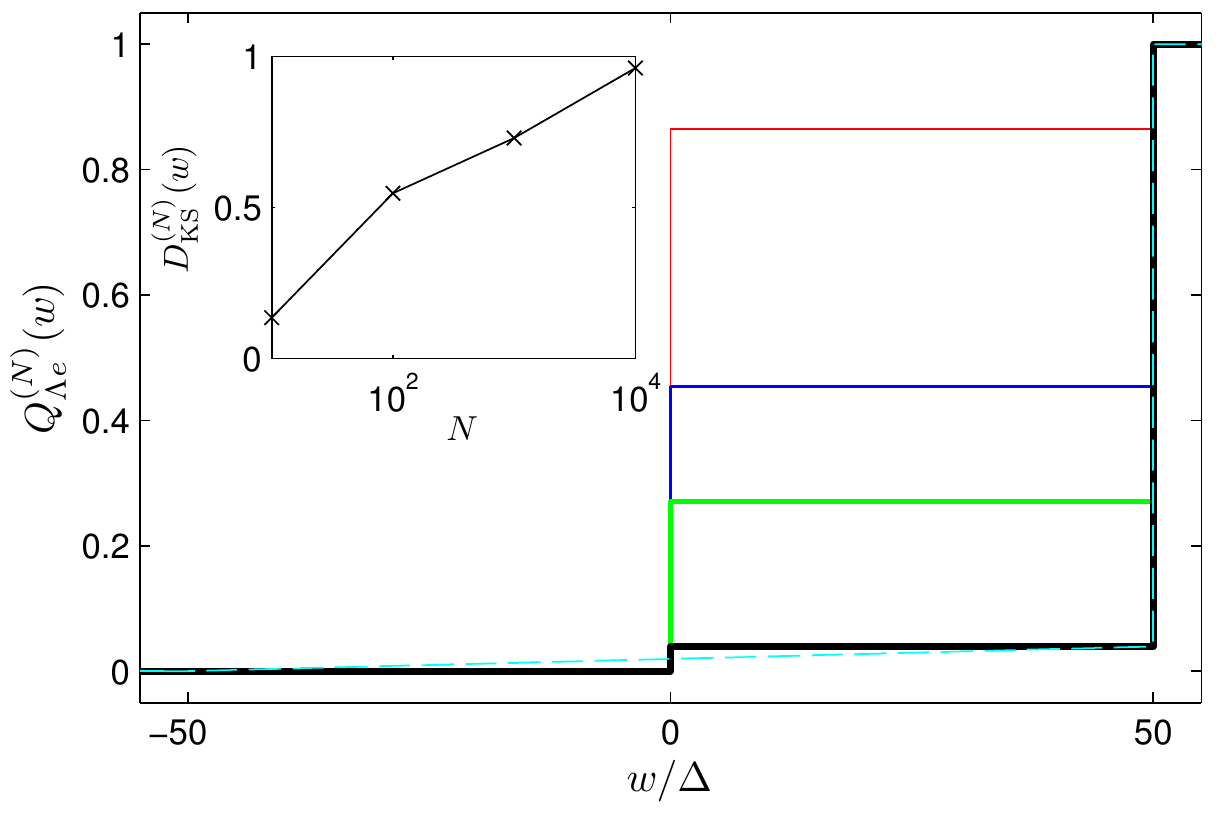}
\caption{Cumulative probability of work from TEMA during a protocol for the LZ problem that is interrupted by $N$ measurements of $\sigma_z$. The system sweep parameters and legends are as in figure \ref{fig:par1betahighzenoapp}, in particular, the color code indicates the same number of power measurements $N$. The initial temperature is small $\beta E_{\mathrm{max}} = 10$. For comparison the cumulative probability computed from $N = 10^4$ measurements of power is also shown (cyan dashed line). The inset shows the Kolmogorov-Smirnov distance between the TEMA and power-based work estimate as a function of $N$.}
\label{fig:par1betahighTEMApower}
\end{figure}

\begin{figure}
\centering 
\includegraphics[width=0.6\textwidth]{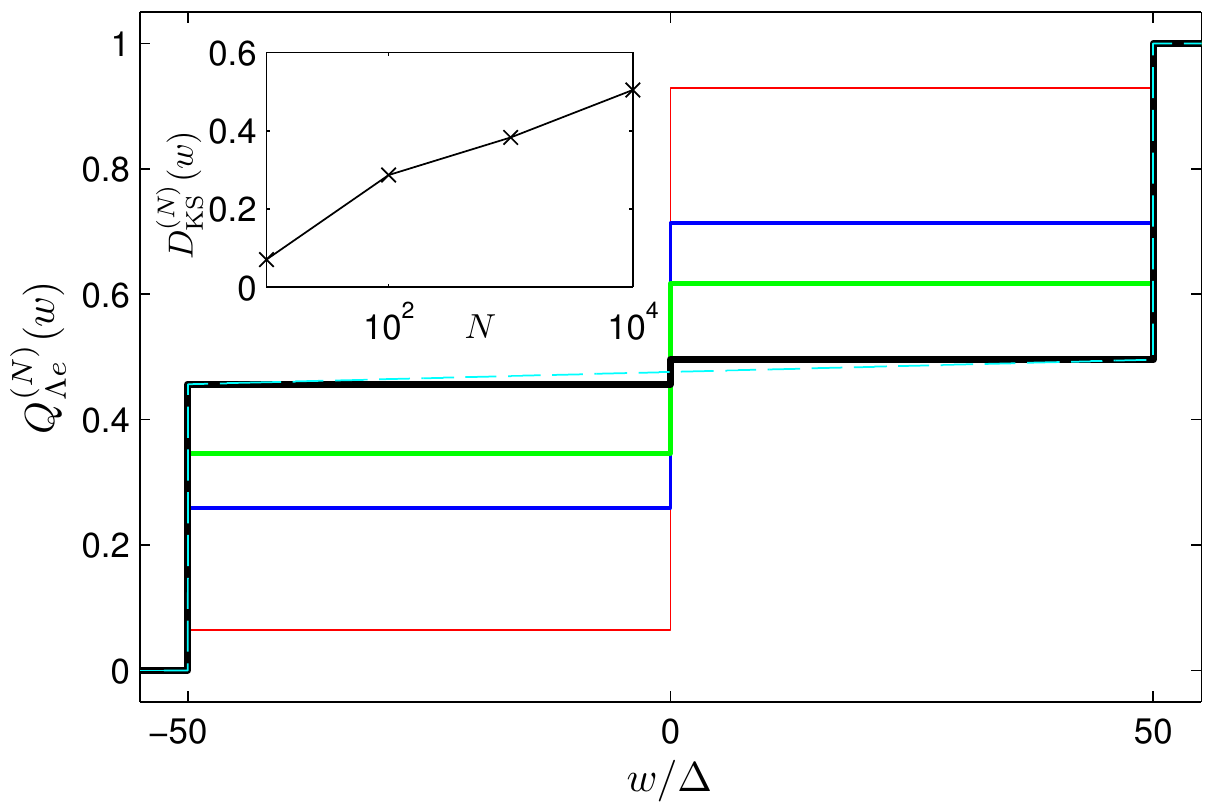}
\caption{Cumulative probability of work from TEMA during a LZ sweep that is interrupted with power measurements. System parameters (and legend) are as in figure \ref{fig:par1betahighTEMApower}. The initial temperature is large $\beta E_{\mathrm{max}} = 0.1$. }
\label{fig:par1betalowTEMApower}
\end{figure}
Figures \ref{fig:par1betahighzenoapp} and \ref{fig:par1betalowzenoapp} display the cumulative probabilities of work computed for different numbers of power measurements $N$ and for two different temperatures. 
The cumulative probability $Q(w) = \int_{-\infty}^w dw' p(w)$ quantifies the probability to find a work value $w' \leq w$ based on the pdf $p(w)$. As a quantitative measure between two work distributions we use the Kolmogorov-Smirnov (KS) distance $D_{KS}$, corresponding to the maximum absolute difference between the respective cumulative probabilities, i.e. $D_{KS}= \max_w |Q_1(w) -Q_2(w)|$. For the parameter values chosen (see figure caption) the  initial eigenstates of the LZ system have significant overlaps with the $\sigma_z$ eigenstates, i.e., $\ket{e_{\mp}(-\tau/2)}_{LZ} \approx \ket{z_{\pm}}$ (the minus sign on l.h.s stands for the ground state). As a result, in the low temperature example in figure \ref{fig:par1betahighzenoapp} one can see that as $N$ is increased, in agreement with (\ref{eq:LZinfpowpdf}), the largest jump of the cumulative probability occurs near $w \approx v \tau/2$. 
Only a small jump at $w \approx -v\tau/2$ is visible due to the low initial occupation of the excited state. For the high temperature example in figure \ref{fig:par1betalowzenoapp}, two jumps of comparable height appear in the large $N$ limit as both eigenstates are occupied in the initial distribution. The cumulative probability from the TEMA is also plotted for reference and it is apparently quite different from the pdf for the power-based work estimate. The largest Kolmogorov-Smirnov distance between the TEMA  and the power-based work distributions results in the Zeno limit.

In earlier work \cite{Campisi11}, it has been shown that though fluctuation theorems for TEMA work are robust to measurements during the protocol, the work statistics itself can be strongly modified. In this light it is interesting to compare the pdf for the power-based work estimate  $p_{\Lambda p}^{(N)}(w)$ with the TEMA work pdf, $p_{\Lambda e}^{(N)}$, in the presence of ($N$) measurements of power during the protocol. Note that while computing $p_{\Lambda e}^{(N)}$, we sum over all possible results of the intermediate power measurements. In figures \ref{fig:par1betahighTEMApower} (low initial temperature) and \ref{fig:par1betalowTEMApower} (high initial temperature), the cumulative probability for the TEMA work pdf with varying number $N$ intermediate power measurements are displayed for the LZ problem. For comparison the cumulative probability for the power-based work estimate with $N=10^4$ measurements is also plotted. In a qualitative way, the two distributions approach each other in the limit of $N \rightarrow \infty$, but the KS distance (shown in the insets) increases with $N$. This apparent contradiction can be resolved by comparing the $N \rightarrow \infty$ limit of the TEMA work pdf with intermediate power measurements for the LZ problem given by
\begin{eqnarray}
p_{\Lambda e}^{\infty}(w) &=& \frac{1}{2}\sqlr{1+\left(\frac{v \tau}{2E_{\mathrm{max}}}\right)^2}\sqlr{p_g \delta(w-E_{\mathrm{max}}) + p_e \delta(w+E_{\mathrm{max}})}  \label{eq:LZintmeaszenopdf}\\
&+& \frac{1}{2}\sqlr{1-\left(\frac{v \tau}{2E_{\mathrm{max}}}\right)^2} \delta(w) \nonumber
\end{eqnarray}
with the equivalent expression for the power-based estimate (\ref{eq:LZinfpowpdf}). In the low temperature case (hence $p_g \approx 1$) depicted in figure \ref{fig:par1betahighTEMApower}, the KS distance is maximised at large $N$,  because the locations at which the cumulative probabilities perform the largest jumps differ. For the power-based estimate, from (\ref{eq:LZinfpowpdf}), the jump occurs at $w = v \tau/2$ whereas for the TEMA based work the jump occurs at $w = E_{\mathrm{max}} = 2\sqrt{(v\tau/4)^2+\Delta^2}$. Hence even for $N \rightarrow \infty$, only in the limit that $\Delta/(v \tau) \rightarrow 0$,  the power-based work and the one from TEMA with intermediate measurements agree. Note that in the limit $\Delta/(v \tau) \rightarrow 0$,  the power operator always commutes with the hamiltonian.

\section{Joint statistics of work from TEMA and power measurements} \label{sec:jointstat}
To further elucidate the differences between the TEMA and power-based work estimates, we consider a modified set-up that allows the simultaneous study of both approaches. In order to combine these two approaches, we consider a thought experiment where in addition to $N$ power-measurements of the type described in the previous section, also energy measurements at the beginning and the end of the force protocol are performed according to the TEMA scheme. The outcome of energies $e_n(0)$ and $e_m(\tau)$ at the beginning and the end of the force protocol and of a sequence $\mathbf{x} = (x_{\alpha_1}, x_{\alpha_2}, \cdots, x_{\alpha_N})$ of eigenvalues of the generalized coordinate $\hX$ at the equally spaced times of measurement $t_1,t_2,\cdots, t_N$  occurs with the joint probability $\mathscr{P}_\Lambda(m,\bx,n)$ given by
\begin{eqnarray}
\mathscr{P}_\Lambda(m,\bx,n) &=  \Tr \sqlr{\Pi_m(\tau)V_{\Lambda}(\bx)\Pi_n(0)\rho_0 \Pi_n(0)V_{\Lambda}^{\dagger}(\bx)} \nonumber\\
&=\abs{\bra{m;\tau}U_{N+1}\ket{\varphi_{\alpha_N}}}^2\abs{\bra{\varphi_{\alpha_1}}U_{1}\ket{n;0}}^2 \nonumber\\
&\quad \times \prod_{k=1}^{N-1} \left | \bra{\varphi_{\alpha_{k+1}}} U_{k+1} \ket{\varphi_{\alpha_{k}}}\right |^2\bra{n;0}\rho_0\ket{n;0} 
\label{eq:jointprobindx},
\end{eqnarray}
where, as before, $\Pi_n(t)$ denotes the projector on the eigenstate $\ket{n;t}$\footnote{We do not allow for degeneracy of the energy eigenvalues for the sake of notational simplicity.} of $H[\lambda(t)]$ belonging to the eigenenergy $e_n(t)$, and $V_\Lambda$ is defined in (\ref{eq:jointprobproj}).   

\begin{figure}
\centering 
\includegraphics[width=0.6\textwidth]{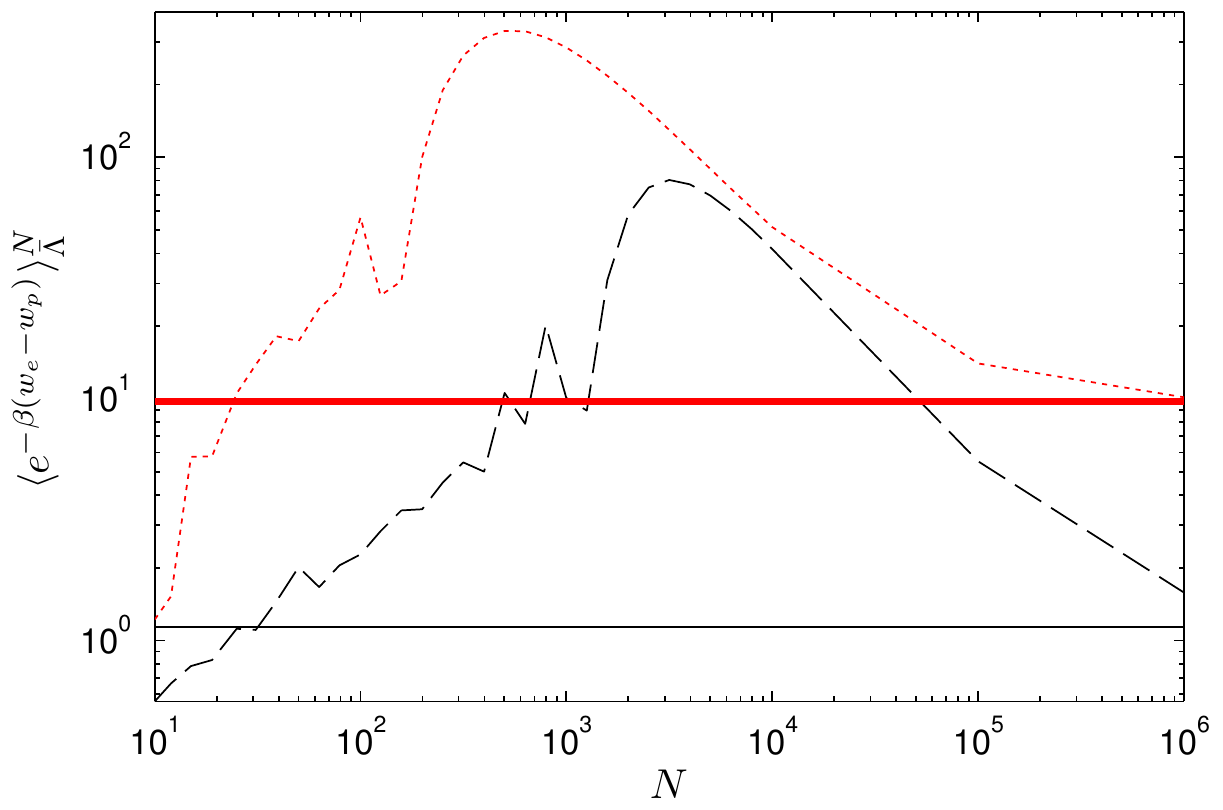}
\caption{Correction factor $\langle e^{-\beta(w_e-w_p)}\rangle^{N}_{\bar{\Lambda}}$ in the integral fluctuation theorem (\ref{eq:jarwpow}) as a function of the number $N$ of power measurements for the LZ problem. The red dotted curve is for the parameters $v = 5 \Delta^2/\hbar$ and the black dashed line is for a faster sweep rate of $v = 40 \Delta^2/\hbar$ with $\tau = 20 \Delta/\hbar$. In both cases the initial temperature is low and satisfies $\beta E_{\mathrm{max}} = 10$. The solid red (thick) and black (thin) lines represent the correction factor computed in the $N\rightarrow \infty$ limit for the slow and fast sweep respectively.}
\label{fig:corrfacjp}
\end{figure}
\begin{figure}
\centering 
\includegraphics[width=0.6\textwidth]{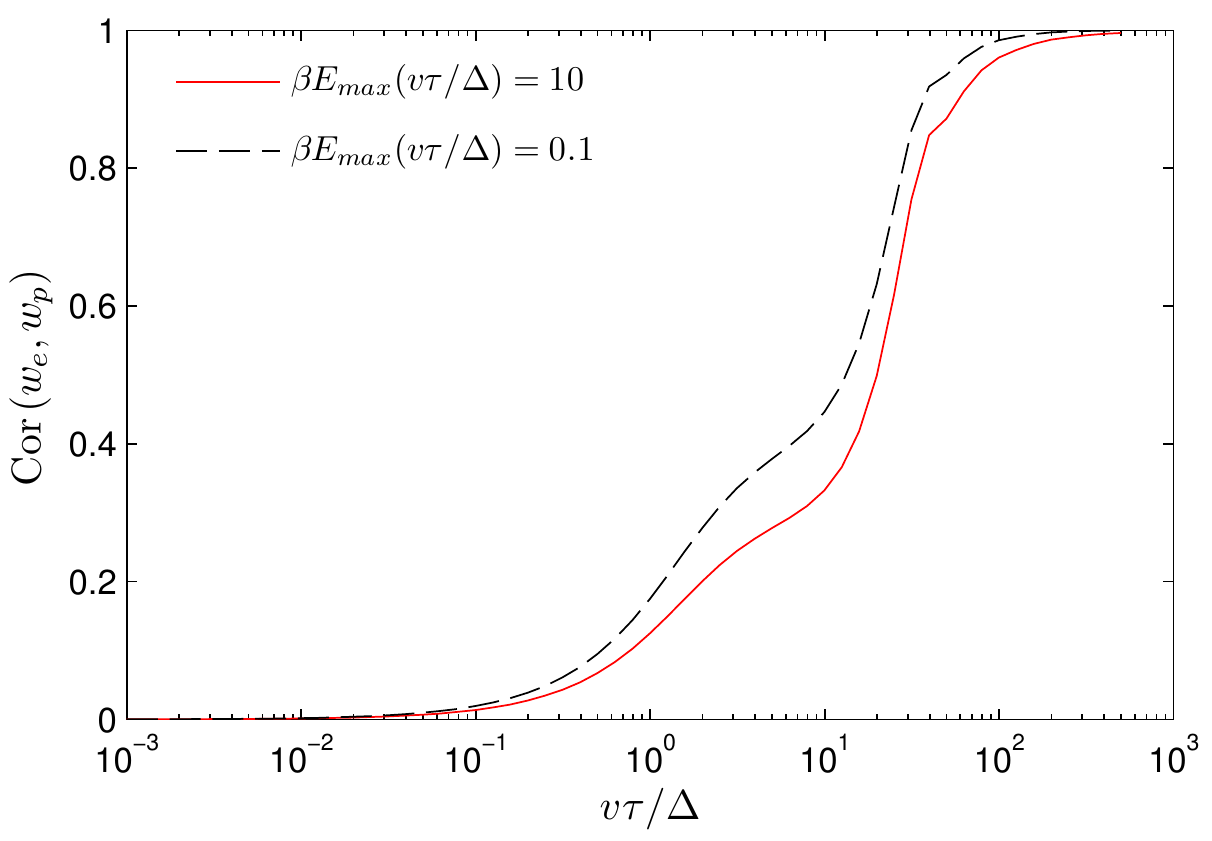}
\caption{Correlation between the TEMA work $w_e$ and power-based estimate $w_p$ for the LZ system is displayed as a function of the sweep velocity $v\tau/\Delta$. The sweep time is fixed at $\tau=20\hbar/\Delta$. The initial temperature, $\beta E_{\mathrm{max}}(v\tau/\Delta)=10$(low temperature, red solid) and $\beta E_{\mathrm{max}}(v\tau/\Delta)=0.1$ (high temperature, black dashed line), is scaled to ensure the same population distribution at different $v$.}
\label{fig:corwewpfnv}
\end{figure}
This joint probability and the according probability for the time-reversed process, $\mathscr{P}_{\bar\Lambda}(m,\bar{\bx},n)$, satisfy a detailed balance-like relation of the form
\begin{equation}
\mathscr{P}_\Lambda(m,\bx,n)= e^{-\beta(\Delta F +e_n(0) -e_m(\tau))}\mathscr{P}_{\bar\Lambda}(n,\bar{\bx},m),
\label{dbr}
\end{equation}
where $\bar{x} = \parlr{x_N,x_{N-1}, \cdots, x_1}$. It holds under two conditions. First, the Hamiltonian $H_0$ and the generalized coordinate $\hX$ must be time reversal invariant meaning that both operators commute with the anti-unitary time reversal operator $\Theta$, $[H_0,\Theta]=[\hX,\Theta]=0$.\footnote{This condition can be modified for Hamiltonians depending on fields changing their parity under time reversal. For details see e.g. \cite{TMYH,AG}} This property allows one to express the time evolution running in the backward time direction in terms of a propagator proceeding along the forward arrow of time \cite{Campisireverr11,AG} in the following way:
\begin{eqnarray}
U^\dagger_{t,s}(\Lambda) = \Theta^\dagger U_{\tau-s,\tau-t}(\bar{\Lambda}) \Theta,
\label{tr}
\end{eqnarray}
The second condition requires that the forward and the backward processes start from canonical equilibrium states $e^{-\beta H[\lambda(0)]}/Z(0)$ and  $e^{-\beta H[\lambda(\tau)]}/Z(\tau)$, respectively. For a detailed derivation we refer to \cite{Watanabe14a}. In terms of the conditional probability
\begin{equation}
   \mathscr{P}_\Lambda(m,\bx|n) = \mathscr{P}_\Lambda(m,\bx,n) e^{\beta(e_n(0) -F(0))}
\label{pfcmxn}
\end{equation}
with the free energy $F(0)$ of the initial state and the analogous conditional probability for the backward process
\begin{equation}
   \mathscr{P}_{\bar{\Lambda}}(n,\bar{\bx}|m) = \mathscr{P}_{\bar{\Lambda}}(n,\bar{\bx},m) e^{\beta(e_m(\tau) -F(\tau))},
\label{pbcmxn}
\end{equation}
one obtains the microcanonical, detailed balance-like relation\footnote{If the energy eigenvalues $e_n(t)$ are degenerate the respective multiplicities $d_n(t)$ have to be taken into account as $ \mathscr{P}_\Lambda(m,\bx|n) d_n(0)= \mathscr{P}_{\bar{\Lambda}}(n,\bar{\bx}|m) d_m(\tau)$, \cite{TMYH}.}
\begin{equation}
 \mathscr{P}_\Lambda(m,\bx|n) = \mathscr{P}_{\bar{\Lambda}}(n,\bar{\bx}|m)   
\label{cdbr}
\end{equation}
generalizing the detailed balance-like relation that holds for the forward and backward conditional probabilities of energy in the absence of intermediate measurements \cite{TMYH}. 

The joint pdf $\mathcal{P}^{N}_\Lambda(w_e,w_p)$ for the two work estimates immediately follows as: 
\begin{eqnarray}
\mathcal{P}_{\Lambda}^{N}(w_e,w_p) &= \sum_{n,m,\flrlr{{\alpha_i}}} \delta \parlr{w_e-\parlr{e_m(\tau)-e_n(0)}} \delta \parlr{w_p-\sum_{i=1}^N \dot{\lambda}(t_i)x_{\alpha_i}h} \label{eq:jointprobwewp} \nonumber\\
& \quad \times \mathscr{P}_\Lambda(m,\bx,n). 
\end{eqnarray}
With the according expression for the time-reversed process,
\begin{eqnarray}
\mathcal{P}^N_{\bar{\Lambda}}(w_e,w_p) &= \sum_{n,m,\flrlr{{\alpha_i}}} \delta \parlr{w_e+\parlr{e_m(\tau)-e_n(0)}} \delta \parlr{w_p+\sum_{i=1}^N \dot{\lambda}(t_i)x_{\alpha_i}h} \nonumber\\
& \quad \times \mathscr{P}_{\bar{\Lambda}}(n,\bar{\bx},m), 
\label{eq:jointprobwewptimerev}
\end{eqnarray}
in combination with the detailed balance-like relation (\ref{dbr}) we find a Crooks-type fluctuation theorem for the joint distribution of TEMA and power-based work:
\begin{eqnarray}
\mathcal{P}^N_{\bar{\Lambda}}(-w_e,-w_p) = e^{-\beta \parlr{w_e - \Delta F}}\mathcal{P}^N_{\Lambda}(w_e,w_p) \label{eq:crooksjointprob}.
\end{eqnarray}
In full agreement with the earlier observation that any number of intermediate projective measurements leaves the Crooks relation, and, consequently, the Jarzynski equality, unchanged while modifying the work statistics \cite{Watanabe14a,Campisi11}. One finds for the marginal distribution $ \mathcal{P}^N_{\Lambda,e}(w_e) = \int dw_p \mathcal{P}^N_{\Lambda}(w_e,w_p)$ for the TEMA work
\begin{equation}
\mathcal{P}^N_{\Lambda,e}(w_e) = e^{-\beta(\Delta F -w_e)} \mathcal{P}^N_{\bar{\Lambda},e}(-w_e).
\label{CN}
\end{equation}
i.e.\ the Crooks relation being fulfilled. Note that the presence of the unnoticed power measurements do have their impact rendering $p_{\Lambda e}(w)$ as defined in (\ref{pe}) different from $\mathcal{P}^N_{\Lambda,e}(w_e)$. Yet both work distributions satisfy the Crooks relation in agreement with \cite{CTH10,Campisi11,Watanabe14a}. 

For the marginal power-based work pdf $ \mathcal{P}^N_{\Lambda,p}(w_p) =\int dw_e  \mathcal{P}^N_{\Lambda}(w_e,w_p)$ only modified fluctuation theorems can be obtained. Integrating (\ref{eq:crooksjointprob}) over all values of $w_e$ one obtains 
\begin{equation}
e^{-\beta(\Delta F - w_p)}  \mathcal{P}^N_{\bar{\Lambda},p}(-w_p) = \mathcal{P}^N_{\Lambda,p}(w_p)  \langle e^{-\beta (w_e - w_p)}|w_p \rangle^N_\Lambda ,
\label{mCp}
\end{equation}
where $\langle \cdot |w_p \rangle^N_\Lambda = \int dw_e \cdot \mathcal{P}^N_{\Lambda}(w_e,w_p)/ \mathcal{P}^N_{\Lambda,p}(w_p)$ denotes a conditional average. This conditional average of the exponentiated work difference determines the modification relative to the strict Crooks relation. In general it depends on the force protocol.  

Changing $\Lambda$ into $\bar{\Lambda}$ and performing an integration over all power-based work values, one obtains 
an integral fluctuation theorem in the form of a modified Jarzynski equality reading
\begin{eqnarray}
\langle e^{-\beta w_p} \rangle^N_{\Lambda} = e^{-\beta \Delta F} \langle e^{-\beta\parlr{w_e-w_p}}\rangle^N_{\bar{\Lambda}}, \label{eq:jarwpow}
\end{eqnarray}
where $\langle \,\,\cdot \,\, \rangle^N_{\Lambda} = \int dw_e dw_p \cdot \mathcal{P}^N_{\Lambda}(w_e,w_p)$ denotes the average over the joint pdf with $N$ power measurements. The correction factor is now determined by the full average of the exponentiated difference between TEMA and power-based work. In general, it is also a protocol dependent factor. In the limit of a large number of power measurements, we can use the approach in section \ref{sec:powmeas} and in a straightforward manner show
\begin{equation}
\langle e^{-\beta\parlr{w_e-w_p}}\rangle^{\infty}_{\bar{\Lambda}} = \frac{\Tr e^{-\beta \sqlr{\lambda(\tau)-\lambda(0)}\hX}e^{-\beta H\sqlr{\lambda(0)}}}{Z(\tau)} \label{eq:jointprobcorrfacinf}.
\end{equation} 
Hence as we remarked in (\ref{Wec}) if $[H_0,\hX] = 0$ is satisfied, the two estimates of work $w_e$ and $w_p$ coincide in the limit of a large number of power measurements (Zeno limit) and the correction factor is unity. Figure \ref{fig:corrfacjp} depicts this correction factor for the LZ problem for two speeds of sweeping (see figure captions for details). The correction factor behaves non-monotonically with $N$ and can be greater or less than $1$ for small $N$. For large $N$ it tends to the correct limiting value as shown in the figure. Moreover for the diabatic sweep $v = 40 \Delta/\hbar$, the correction factor is in general significantly smaller than the moderate sweep rate case of $v = 5 \Delta/\hbar$. This can be anticipated since in the diabatic sweep case with large $v$, except for a small interval around $t=0$ the total LZ Hamiltonian (\ref{eq:LZmodel}) approximately commutes with the power operator as the diagonal terms $ \pm v t/2$ dominate the off-diagonal coupling $\Delta$. 

So far in this section we elucidated some detailed and integral fluctuation relations satisfied by joint TEMA and power-based work estimates and their respective marginals and also pointed to formal differences between these two estimates. One quantitative measure of the relation between the two estimates of work is the mutual information $\Delta I_{\Lambda}(w_e:w_p) \equiv \ln \frac{\mathcal{P}_{\Lambda}^N(w_e,w_p)}{\mathcal{P}_{\Lambda,e}^N(w_e)\mathcal{P}_{\Lambda,p}^N(w_p)}$. While this measure locally quantifies the interdependence of $w_e$ and $w_p$, we consider, the correlation function, as a global measure. It is defined as:
\begin{eqnarray}
\mathrm{Cor}(w_e,w_p) = \frac{\sigma_{w_e,w_p}}{\sigma_{w_e}\sigma_{w_p}} \label{eq:corrdefn},
\end{eqnarray}
where the covariance and standard deviation of the two variable are given by
\begin{eqnarray*}
\sigma_{w_e,w_p} &=& \langle w_e w_p \rangle^N_{\Lambda} - \langle w_e\rangle_{\Lambda}\langle w_p\rangle^N_{\Lambda},\\ 
\sigma_{w_{j}}^2 &=& \langle w_{j}^2\rangle^N_{\Lambda} - \parlr{\langle w_j\rangle^{N}_{\Lambda}}^2 \,\, ;j=p,e.
\end{eqnarray*}
Here, the averages are performed over the joint pdf given by (\ref{eq:jointprobwewp}). The formal expressions for the correlation in general are not very transparent and instead it is more illuminating to consider a specific example. We choose the LZ problem introduced in section \ref{sec:powmeas} and fix the number of power measurements at $N=100$. With a fixed time of sweep at $\tau= 20\hbar/\Delta$, we ask how the correlation between the two estimates of work varies as the velocity of sweep $v$ is changed. In figure \ref{fig:corwewpfnv}, we present our results for two thermal initial states. For each sweep rate the temperature is chosen such that $\beta E_{\mathrm{max}}(v,\Delta) = \{10,0.1\}$ corresponding to fixed populations of the ground state independent of the sweep rate. In figure \ref{fig:corwewpfnv} we can clearly see that as $v \tau/\Delta$ is increased, the two estimates become more and more correlated. This can again be understood as the effect of making the power term $\frac{vt}{2} \sigma_z$ much larger in magnitude compared to the time-independent part $\Delta \sigma_x$. This effectively renders the commutator between $[H_0,\hX] \sim 0$ for most of the interval and hence the power-based work estimate agrees well with the TEMA based one. Secondly we also see that at the smaller value of $\beta E_{\mathrm{max}}$ corresponding to a larger temperature, since the system is more ``classical", the correlation between the two estimates is better.

In the next section we consider weak continuous measurements of power to estimate the work as opposed to projective measurements considered thus far in the paper.

\section{Weak continuous measurement of power} \label{sec:fintime}
In the previous sections we modelled the monitoring of the supplied power by means of projective measurements of the generalised coordinate $\hX$. As we have shown, in the limit of large number of such measurements, the system dynamics is frozen in the basis of the generalized coordinate and the unitary dynamics generated by the driving plays no role. The work statistics in this limit also differs from the ones determined by the TEMA. In an attempt to mitigate this situation we consider a {\it weak}  continuous  measurement of the generalised coordinate and determine power from such a measurement. One might hope that in this case although the estimate of work will be affected by errors inherent in a weak measurement process, the measurement backaction will not be so overwhelming as to render the unitary dynamics moot. To this end in what follows we will use the theory of continuous quantum measurement developed in \cite{Contmeaspaps} for our specific situation of determining work statistics from power measurements.

Before considering a particular scenario of continuous coordinate measurements in more detail, we shortly discuss the measurement of the relevant coordinate by means of Gaussian Kraus operators \cite{Sudarshankraus} as a particular example of a weak instantaneous measurement. More general Kraus operators generating a positive operator valued measure (POVM) of the coordinate are discussed in the appendix \ref{app:genpower}.

\subsection{Instantaneous Gaussian coordinate measurements}
We now assume that the measurement of the generalized coordinate $\hX$ is performed with measurement operators consisting of weighted sums of eigen-projection operators rather than of a single one. Choosing Gaussian weights we have
\begin{eqnarray}
M_{\alpha} &=& \frac{1}{\parlr{2 \pi \sigma^2}^{1/4}}\exp \sqlr{-\frac{(\alpha - \hX)^2}{4\sigma^2}} \label{eq:Gaussianmeas}\\
&=& \frac{1}{\parlr{2 \pi \sigma^2}^{1/4}}\sum_{n}\exp \sqlr{-\frac{\parlr{\alpha - x_n}^2}{4\sigma^2}} \Pi_{n}^{X},\nonumber
\end{eqnarray}
where $\alpha$ denotes the pointer state indicating the measured value of the coordinate,(Note that this choice is slightly different from the one adopted in Appendix \ref{app:genpower}, where we assume that the set of pointer states consists only of the eigenvalues of the generalized coordinate in contrast to the continuous range of $\alpha$ values in (\ref{eq:Gaussianmeas})) and $\sigma^2$ the variance of the error distribution of measured coordinate values. The work that can be estimated from $N$ generalised measurements of $\hX$ equally spaced in time, is given by $w_p = \sum_{n} h \dot{\lambda}\parlr{t_n} \alpha_n$ and its pdf  takes the form (see appendix \ref{app:genpower} for details)
\begin{eqnarray}
p_{\Lambda p}^{(N)} (w)=\displaystyle \int {\prod_{k=1}^N d \alpha_k} \,\, \delta \parlr{w-\sum_{k} h \dot{\lambda}\parlr{t_k} \alpha_k } \Tr \mathcal{M}^\dagger \mathcal{M} \rho_0 \label{eq:gaussmeaspdf},
\end{eqnarray}
where $\mathcal{M} = M_{\alpha_{N}}(t_N) M_{\alpha_{N-1}}(t_{N-1}) \cdots M_{\alpha_{1}}(t_1)$. In the above, we have used the Heisenberg picture representation of the Gaussian measurement operators introduced in the appendix (\ref{eq:jointprobformgenheisform}). A natural way to define continuous measurements in this framework is to consider weak measurements characterized by a variance that increases proportionally to the inverse of the time $h$ between two measurements, i.e. as $\sigma^2 = 1/(8\kappa h)$ where $\kappa>0$ quantifies the measurement strength.  Hence, with an increasing number of measurements less and less information is gained from a single measurement. When we take such a scaling for the variance, the measurement operator (\ref{eq:Gaussianmeas}) can be expanded as
\begin{eqnarray}
M_{\alpha} \propto \exp \sqlr{-2\kappa h (\alpha-\hX)^2} = I -2\kappa h (\alpha-\hX)^2 + O(h^2) \label{eq:gaussmeasopexpand}.  
\end{eqnarray}
In appendix \ref{app:genpower} we take the generalised measurement operators to be independent of the time step. Since the unitary operators can always be written as $U_k = I + O(h)$, to lowest order in $h$ we obtain an expression for the pdf (\ref{eq:genmeaspower}) that depends only on the measurement operators and the initial state of the system, provided that the time-independent part of the Hamiltonian, $H_0$, is symmetric in the $\hX$ basis. In contrast, for the weak measurement case (\ref{eq:gaussmeasopexpand}) implies that both the measurement and unitary operators enter (\ref{eq:gaussmeaspdf}) at the same order in $h$ and the system's state is unchanged to the lowest order. In this situation to the best of our attempts, we are not able to determine a simple expression for the pdf (\ref{eq:gaussmeaspdf}) in the $h \rightarrow 0$ limit. In the next subsection we follow another path to tackle this problem by focusing on the differential equation that describes a system that is under continuous measurement and is simultaneously driven by a change in the Hamiltonian. For the sake of completeness we note that although we cannot find a simple closed form expression for the pdf (\ref{eq:gaussmeaspdf}) in general, for the trivial case of the system Hamiltonian commuting with the measured coordinate $[H_0,\hX] = 0$, we can easily compute the pdf as
\begin{eqnarray}
p_{\Lambda p}^{(\infty)} (w) &=& \frac{4 \kappa}{\pi \int_0^{\tau} dt \dot{\lambda}(t)^2} \nonumber\\
& \Tr &\sqlr{\exp \flrlr{-\frac{4 \kappa}{\int_0^{\tau} dt \dot{\lambda}(t)^2} \parlr{w - \sqlr{\lambda(\tau) - \lambda(0)} \hX}^2} \rho_0} \label{eq:gaussmeasinf}.
\end{eqnarray}
The only difference to the power-based work pdf for projective measurements, (\ref{eq:projinfmeaspowpdf}) is the replacement of the delta-function by a Gaussian weight under the trace with a variance depending on the force protocol. The average of the exponentiated work yields the expression
\begin{equation}
\langle e^{-\beta w_p} \rangle = e^{\frac{\beta^2}{16 \kappa } \int_0^\tau dt \dot{\lambda}^2(t)} \Tr e^{-\beta\sqlr{\lambda(\tau)-\lambda(0)}\hX} \rho_0 = e^{\frac{\beta^2}{16 \kappa } \int_0^\tau dt \dot{\lambda}^2(t)} e^{-\beta \Delta F},
\label{ebw}
\end{equation}
where the second equality on the the right-hand side follows from the commutation of $H_0$ and $\hX$. Thus we find that even for the trivial situation with $[H_0,\hX] = 0$ work defined by the integral of weak continuous measurement of power does not satisfy the Jarzynski relation.

\begin{figure} 
\centering
\includegraphics[width=0.6\textwidth]{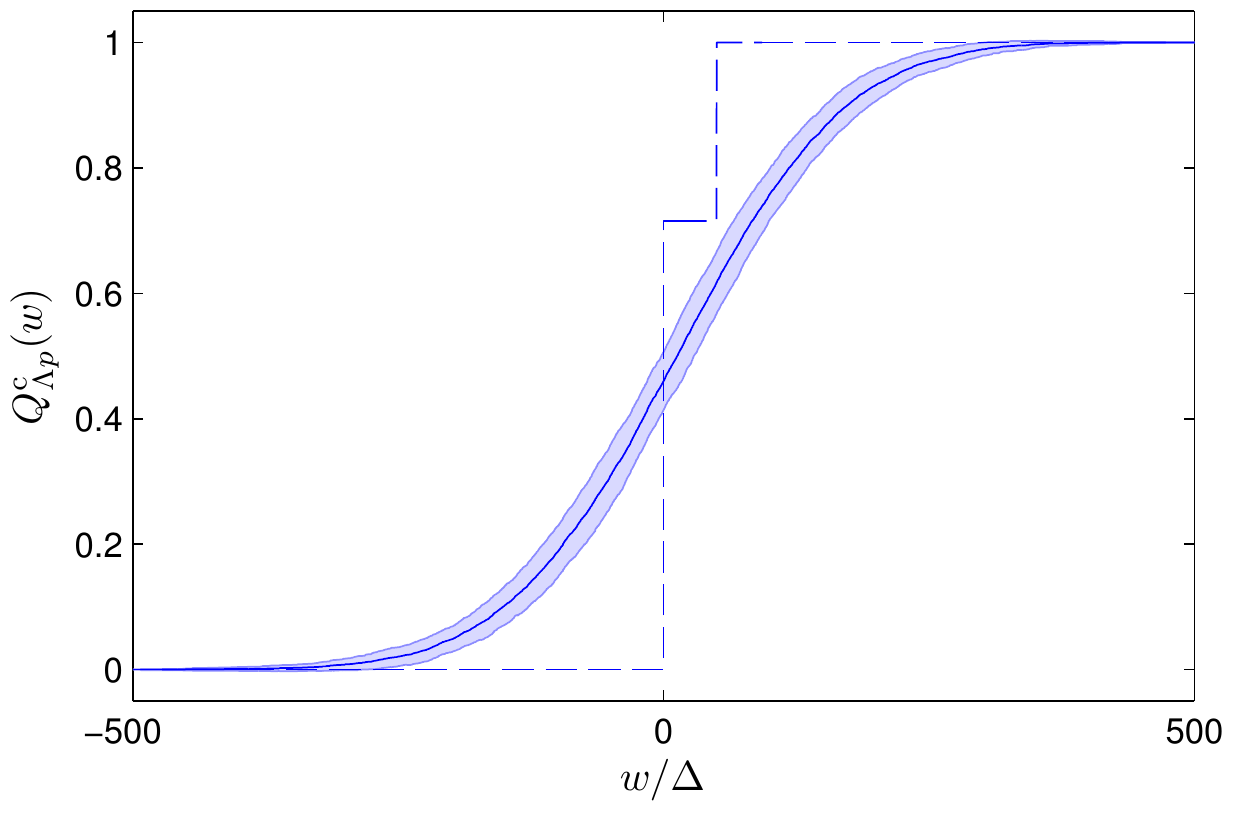}
\caption{Cumulative probability estimated from continuous measurement of power in the LZ problem with $v = 5 \Delta^2/\hbar$, $\tau = 20 \hbar/\Delta$ (blue solid line) and measurement strength $\kappa = 0.001 \Delta/\hbar$. The initial temperature is small $\beta E_{\mathrm{max}} = 10$. The results were obtained from 10000-trajectory simulations of the SME (\ref{eq:contsme}). The shaded region represents the error in the computed pdf represented by the solid line (see text). The dashed blue line represents the pdf computed by TEMA.}
\label{fig:paradbbetahighkapsmLZ}
\end{figure}

\begin{figure}
\centering 
\includegraphics[width=0.6\textwidth]{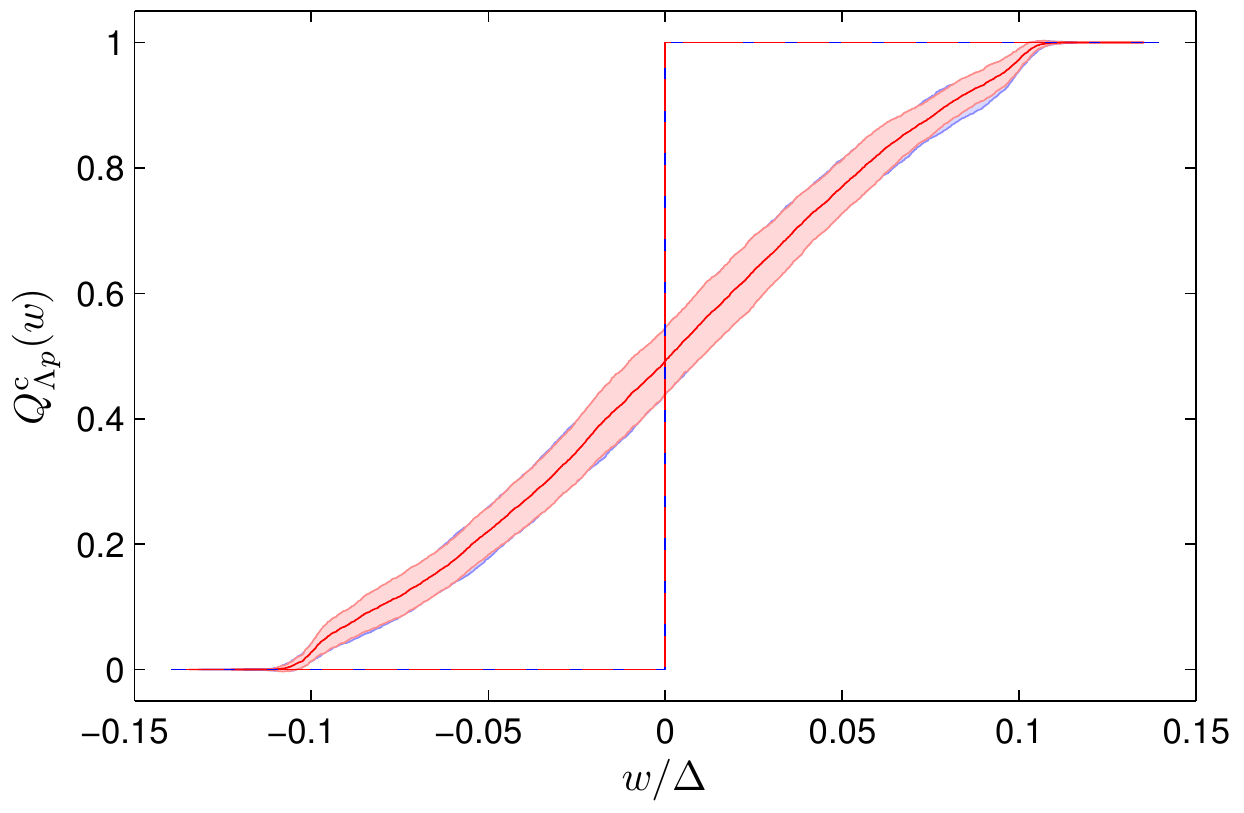}
\caption{Cumulative probability of work estimated from continuous measurement of power in the LZ problem with a large measurement strength $\kappa = 4 \Delta/\hbar$. The system parameters are chosen as $v =  0.01 \Delta^2/\hbar$, $\tau = 20 \hbar/\Delta$ to correspond to an adiabatic LZ sweep. The initial temperature is large $\beta E_{\mathrm{max}} = 0.1$ for the red curve and smaller $\beta E_{\mathrm{max}} = 10$ for the blue curve.}
\label{fig:paradbkapzeno}
\end{figure}

\begin{figure}
\centering 
\includegraphics[width=0.6\textwidth]{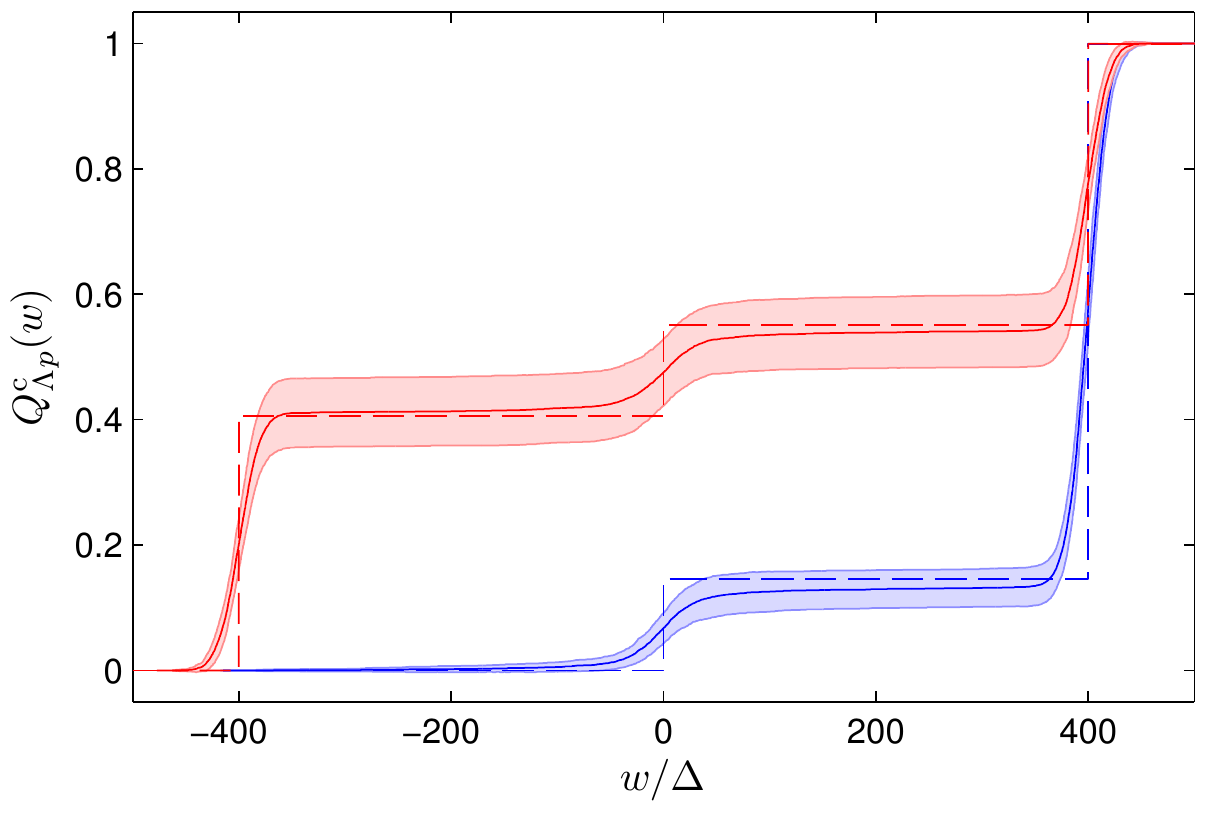}
\caption{Cumulative probability of work estimated from continuous measurement of power in the LZ problem with a large measurement strength $\kappa = 4 \Delta/\hbar$. The system parameters are chosen as $v =  40 \Delta^2/\hbar$, $\tau = 20 \hbar/\Delta$ to correspond to a diabatic LZ sweep. The initial temperature is large $\beta E_{\mathrm{max}} = 0.1$ for the red curve and smaller $\beta E_{\mathrm{max}} = 10$ for the blue curve.}
\label{fig:pardiabkapzeno}
\end{figure}
\subsection{Continuous measurements}
In a final attempt we shall relax the assumption inherent in both the projective and the generalized measurement approach that a measurement consists in an instantaneous event that interrupts the unitary dynamics of the system. For that purpose we adopt a model proposed in \cite{Jacobs06}. The time evolution of the density matrix caused by the unitary dynamics in the presence of continually performed measurements of an observable $\hX$ is described by the following non-linear stochastic master equation (SME)
\begin{eqnarray}
\dot{\rho}(t) &=& -\frac{i}{\hbar} \sqlr{H [\lambda(t)],\rho(t)}  - \kappa [\hX,[\hX,\rho(t)]]  \label{eq:contsme}\\
 &+&\sqrt{2\kappa}\parlr{\hX \rho(t) + \rho(t) \hX - 2 \langle \hX(t) \rangle \rho(t)} \xi(t) \nonumber,
\end{eqnarray} 
where  $\xi(t)$ denotes Gaussian white noise of unit intensity, i.e. $\langle \xi(t) \xi(s) \rangle = \delta(t-s)$. Apart from the first, Hamiltonian term on the right-hand side, the second and third terms are supposed to describe the impact of continuous measurements of the observable on the system's dynamics. The second term models the average influence of the measurements, and the third, nonlinear and random term accounts for the influence of an individual run of measurements. The meanfield-like non-linearity  $2 \langle \hX \rangle \rho$ guaranties the conservation of the normalization of the density matrix in the presence of the measurement-induced fluctuations. The average $\langle \hX \rangle$ is performed with respect to the fluctuating density matrix $\rho$ and therefore it is itself a random quantity. In appendix \ref{app:smederiv}, for the sake of pedagogy, we reproduce a derivation of (\ref{eq:contsme}) following \cite{Jacobs06}. There we utilise Gaussian measurement operators with variances that are scaled inversely with the time step like earlier in this section.

According to the theory of continuous measurements \cite{Jacobs06}, the result $\alpha(t)$ of the measurement of $\hX$ at the instant of time $t$ is given by
\begin{eqnarray}
\alpha(t) = \langle \hX (t) \rangle + \frac{1}{\sqrt{8\kappa}} \xi(t) \label{eq:discoutcome},
\end{eqnarray}
and consequently, the work based on the measurement of the power becomes
\begin{eqnarray}
w_p^{\mathrm{c}} &= \int_0^{\tau} dt \dot{\lambda}(t) \alpha(t) \nonumber \\
&=  \int_0^{\tau} \dot{\lambda}(t) \parlr{\langle \hX \rangle(t)  + \frac{1}{\sqrt{8\kappa}}\xi(t)} dt. \label{eq:workcontpow}
\end{eqnarray}  
The SME is understood in the It\={o} sense. Therefore, the third term on the right-hand side of (\ref{eq:contsme}) disappears upon an average over realizations of the Gaussian white noise yielding for the averaged density matrix $\bar{\rho}$ the linear master equation
\begin{eqnarray}
\frac{d \bar{\rho}}{dt} &=& -\frac{i}{\hbar} \sqlr{H[\lambda(t)],\bar{\rho}} - \kappa [\hX,[\hX,\bar{\rho} \,]] \label{eq:contsmeavg},
\end{eqnarray}
It describes the state of the system in a non-selective measurement of the coordinate $\hX$. 

Before discussing the non-linear SME by means of numerical simulations of an example we shortly mention a linear SME that is obtained from (\ref{eq:contsme}) by disregarding  the mean-field-type contribution proportional to the average $\langle \hX \rangle$ \cite{LSE,LSEjacobs}. As a consequence, the normalization of the density matrix is no longer conserved. Even this linear equation is rather complicated for our scenario due to the explicit time dependence of the Hamiltonian. In general we find that it can be solved analytically only if $H_0$ and $\hX$ commute. The resulting work pdf coincides as expected with that of the continuous action of weak instantaneous Gaussian measurements given by (\ref{eq:gaussmeasinf}).  

As an example, we generated $10000$ realizations of the density matrix by solving the SME in (\ref{eq:contsme}), displayed in figures \ref{fig:paradbbetahighkapsmLZ}, \ref{fig:paradbkapzeno} and \ref{fig:pardiabkapzeno} for the LZ model (\ref{eq:LZmodel}) in the presence of continuous measurements of $\sigma_z$. For this purpose we used an implicit stochastic Runge-Kutta scheme of order $3/2$ \cite{KloedenPlaten,Steckbook} and checked for numerical convergence of single trajectories using the method of consistent Brownian paths \cite{Steckbook}. We also checked that, for the chosen time step, the density matrix remains normalised to $1$ to high accuracy at all times. We obtained the work pdf by computing a total of 10000 trajectories. For the LZ problem the work follows from (\ref{eq:workcontpow}) for each trajectory as  
\begin{eqnarray}
w_p^{\mathrm{c}} = \frac{v}{2} \sqlr{\int_{0}^{\tau} dt \langle \hsig_z \rangle(t)  + \frac{1}{\sqrt{8\kappa}} \int_0^{\tau} dt \,\xi(t)}. \label{eq:workdefncontLZ}
\end{eqnarray}
Evidently, the work values in a continuous measurement of $\sigma_z$ span the entire real line due to the additive contribution of the integral of Gaussian white noise. In order to estimate the work pdf from a finite number of simulations, we introduced a binning of the size $\Delta w_p = v h$ with the same time-step $h$ as used in the discretisation of the SME leading to the same set of possible work values as in (\ref{eq:wpLZ}) for $N = \tau/h-1$ measurements. Based on the independence of different trajectory simulations we could estimate the inherent statistical error of the work histograms as the sample variance of 100 blocks of 100 trajectories each.

When the measurement strength $\kappa$ is small such that $\kappa \tau \ll 1$, the work estimate (\ref{eq:workdefncontLZ}) is dominated by the noise term and the resulting pdf is approximately Gaussian distributed with zero mean. This feature is independent of the other system parameters such as $v$ and $\Delta$. In figure \ref{fig:paradbbetahighkapsmLZ} we plot the cumulative probability for the power-based work computed from solving the SME for the same LZ parameters as in section \ref{sec:powmeas}. 

When $\kappa \tau \gtrsim 1$, in the work expression the contribution from $\langle \sigma_z \rangle(t)$ dominates over the noise. Thus the behaviour of the cumulative probability in figures \ref{fig:paradbkapzeno} and \ref{fig:pardiabkapzeno} can be well understood from the behaviour of $\langle \sigma_z \rangle(t)$. For instance, from (\ref{eq:workdefncontLZ}) this immediately explains why the range of allowed work values is comparable to the interval $[-v\tau/2,v\tau/2]$. Also note that for our choice of parameters in figures \ref{fig:paradbkapzeno} and \ref{fig:pardiabkapzeno}, we are also in the strong measurement regime of $\kappa > \Delta/\hbar$ for the LZ problem where the coherent dynamics rate $\Delta$ is trumped by the measurement backaction. It is known \cite{Gagen93,Haikka14,Novelli15} that for the LZ system under continuous measurement,  the behaviour of single trajectory solutions of (\ref{eq:contsme}) goes from near unitary at very small measurement strength to the so-called ``random-telegraph'' dynamics at the strong measurement $\kappa>\Delta/\hbar$ regime (In \cite{Gagen93,Novelli15} the dynamics in the strong measurement regime is referred to as the quantum Zeno effect. In our work we reserve the latter term for the complete freezing of unitary dynamics achieved by repeated \textit{projective} measurements discussed in section \ref{sec:powmeas}). The random-telegraph behaviour is characterized by the population difference of diabatic states $\langle \sigma_z \rangle$ remaining localised either at $\pm 1$ and undergoing rapid transitions between the two values at different times during the evolution (see discussion in Sec. \Rmnum{3} B of \cite{Gagen93}). In the adiabatic sweep case depicted in figure \ref{fig:paradbkapzeno}, since $v \tau \ll \Delta$ the initial state is off-diagonal (with almost equal distribution amongst the $\pm 1$ eigenstates) in the diabatic i.e. $\sigma_z$ basis and throughout the dynamics the system Hamiltonian has a significant off-diagonal component in the diabatic basis i.e. $[H[\lambda(t)],\sigma_z] \neq 0$. Since there is equal contribution from both $\langle \sigma_z \rangle = \pm 1$ to (\ref{eq:workdefncontLZ}), the work pdf is centered about $w = 0$ in figure \ref{fig:paradbkapzeno}. Nonetheless the power-based cumulative probability does not capture the sharp jump at $w=0$ of the TEMA estimate. In the diabatic sweep in figure \ref{fig:pardiabkapzeno}, the system Hamiltonian and the initial state are almost diagonal in the diabatic basis giving cumulative probabilities that have sharp jumps at $w = v \tau/2$ (low initial temperature) or $w = \pm v \tau/2$ (high temperatures). Also in figure \ref{fig:pardiabkapzeno} the jumps in the power-based and TEMA estimates occur at approximately the same values of $w$. We note that this behaviour owes its explanation, as noted in section \ref{sec:powmeas}, to the fact that for our choice  $v \tau/4 \gg \Delta$ the magnitude of maximum work possible in TEMA $E_{\mathrm{max}} \approx v \tau/2$ almost coincides with the location of the jumps in the power measurement cumulative probability. 

In summary we find that the work pdf estimated even by weak continuous measurement of the power does not reproduce the behaviour of the TEMA estimate in general. For very weak measurement strengths $\kappa \tau \ll 1$, the estimate is noise dominated and has Gaussian behaviour irrespective of $v$ and $\Delta$. When $\kappa \tau \gtrsim 1$ (strong measurement regime) the pdf depends on the sweep rate $v$ (for a fixed value of $\tau$) which sets the extent to which the total Hamiltonian is off-diagonal in the measurement basis.

\section{Conclusion} \label{sec:conclusion}
This work serves as a detailed illustration of the difficulties involved in defining work in quantum systems in a manner analogous to classical systems. We considered the statistical properties of work, defined as the integral of supplied power, for a quantum system. We showed that even a careful definition of work in terms of repeated measurements of the system's instantaneous power leads to a statistics quite different from the usual two energy measurement approach (TEMA) for defining work. In the power-based approach in general even the Jarzynski equality or the Crooks relation do not hold. In the limit of a large number of projective measurements of the instantaneous power we found that the quantum Zeno effect leads to a freezing of the system's dynamics in the power operator's basis. The statistics of power-based work in this limit is very different from TEMA except for the trivial case when the power operator commutes with the system Hamiltonian at all times. Furthermore we carried out a detailed comparison of the power-based work and the TEMA work estimate by considering the joint pdf of both types of work estimates in a setting that combines both approaches. We obtained a Crooks type fluctuation theorem for the joint probability distribution and modified integral fluctuation theorems for the marginal power-based work distribution. We also studied the correlation between the two estimates of work using the joint probability for the LZ problem. Finally, relaxing the condition of projective, instantaneous measurements, weak continuous measurement of power was discussed within the formalism of stochastic master equation (SME). Using the Landau-Zener problem as an example, we determined the pdf of power-based work numerically and analysed its properties in the limiting cases of weak and strong measurement strengths. We found that since the power operator in general does not commute with the total Hamiltonian of the system, also within the framework of continuous measurements the power-based estimate is not able to reproduce the qualitative features of the TEMA work estimate.

Finally we would like to note that the treatment of continuous measurements in terms of an SME also leads to a Lindblad-type master equation, see   (\ref{eq:contsmeavg}), similarly as obtained in various recent publications  \cite{Esposito,Horowitz,Horowitz13,Hekking13} which are concerned with the definition of work and heat and the quest for fluctuation theorems in open quantum systems. There, the presence of terms in the equation of motion of the reduced density matrix describing energy-non-conserving effects is caused by the interaction with the environment. In contrast, in the present case it is solely the result of a weak but continuously acting measurement that leads to similar formal structures.    

\ack
We thank Sebastian Deffner for useful discussions and the anonymous referee for useful comments. We acknowledge support by the Max Planck Society, the Korea Ministry of Education, Science and Technology (MEST), Gyeongsangbuk-Do, Pohang City, as well as by the Basic Science Research Program through the National Research Foundation of Korea funded by the MEST (No. 2012R1A1A2008028). P.T. thanks the Foundation for Polish Science (FPS) for providing him with an Alexander von Humboldt Polish Honorary Research Fellowship. G.W. also acknowledges support under Project Code (IBS-R024-D1).

\appendix
\renewcommand\thesection{\Alph{section}}
\section{Generalized power measurements}
\label{app:genpower}
In this appendix we consider generalized measurements of the power observable during the protocol and prove a bound on the exponential average of the work estimate obtained in this manner. We replace the projective measurements of the coordinate $\hX$ in the definition of joint probability (\ref{eq:jointprobproj}) by a set of self-adjoint, i.e. minimally disturbing generalized measurements which have the following form
\begin{eqnarray}
M_{\alpha_j} = \displaystyle \sum_{\alpha_k} \sqrt{p(\alpha_j| \alpha_k)} \,\,\Pi_{\alpha_k}^{X}. \label{eq:genmeasform}
\end{eqnarray} 
We associate with each outcome $\alpha_j$ of the generalized measurement the coordinate eigenvalue $x_{\alpha_j}$. $p(\alpha_j|\alpha_k)$, then denotes the conditional probability for erroneous assignment of eigenvalue $x_{\alpha_j}$ to $x_{\alpha_k}$ by the generalized measurement. These must be real and positive and in accordance with partition of unity, $\sum_{\alpha_j} M_{\alpha_j}^{\dagger} M_{\alpha_j} = I$, they add up to unity:
\begin{eqnarray}
\sum_{\alpha_j} p(\alpha_j|\alpha_k) = 1 \label{eq:normprob}.
\end{eqnarray}
Thus, in the same manner as in section \ref{sec:powmeas} we can write down the pdf for the work estimated by summing up results of generalized measurements of power as:
\begin{eqnarray}
p_{\Lambda p}^{(N)} (w)=\displaystyle \sum_{\flrlr{\alpha_i}} \delta \parlr{w-\sum_{i=1}^{N} \dot{\lambda}(t_i)x_{\alpha_i}h} \Tr V_{\Lambda M}\parlr{\bx} \rho_0 V_{\Lambda M}^{\dagger}\parlr{\bx},\label{eq:genfinmeaspowpdf}
\end{eqnarray}
where
\begin{eqnarray}
V_{\Lambda M}\parlr{\bx} &=& U^{\dagger}_{t_N,t_0} M_{\alpha_{N}} U_N M_{\alpha_{N-1}} \cdots U_2 M_{\alpha_{1}} U_1 \label{eq:jointprobformgen}\\
&=&  M_{\alpha_N}(t_N)M_{\alpha_{N-1}}(t_{N-1})\cdots M_{\alpha_1}(t_1) \label{eq:jointprobformgenheisform},
\end{eqnarray}
where in the second line we have defined Heisenberg picture measurement operators $M_{\alpha_j}(t_j) = U_{t_j,t_0}^{\dagger}M_{\alpha_j}U_{t_j,t_0}$.

In the limit of large $N$, in line with the treatment in section \ref{sec:powmeas}, the unitary operators in (\ref{eq:jointprobformgen}) can be expanded as $U_j = I + O(h)$. Provided the time-independent part of the Hamiltonian $H_0$ is symmetric in the coordinate $\hX$'s basis i.e.\ $\bra{\varphi_{\alpha_j}}H_0\ket{{\varphi_{\alpha_k}}} = \bra{\varphi_{\alpha_k}}H_0\ket{{\varphi_{\alpha_j}}}$, the individual terms that are of order $h$ cancel amongst themselves. This ensures that such terms do not sum up to a term that adds to the leading order contribution. Hence to  leading order in $h$ the work distribution is given by:
\begin{eqnarray}
p_{\Lambda p}^{(N)} (w)&=&\displaystyle \sum_{\flrlr{\alpha_i}} \delta \parlr{w-\sum_{i=1}^{N} \dot{\lambda}(t_i)x_{\alpha_i}h} \Tr M_{\alpha_N}^2 M_{\alpha_{N-1}}^2 \cdots M_{\alpha_1}^2 \rho_0 \nonumber \\ &+& O(h), \label{eq:genmeaspower}
\end{eqnarray}
where we have used the fact that generalized measurement operators at different times commute. In the trivial case when the Hamiltonians at different times commute, the unitary operators $U_k$ commute with the measurement operators and (\ref{eq:genmeaspower}) becomes exactly valid not just when $h \rightarrow 0$.

Unlike for the case of projective measurements of $\hX$, further simplification of the expression (\ref{eq:genmeaspower}) by taking the limit $N \rightarrow \infty$ is not possible without choosing a specific form for the error distribution function $p(\alpha_j|\alpha_k)$. Nonetheless as we show below, we can derive an inequality for the exponential average under some assumptions. The exponential average of the work estimate to zeroth order in $h$ is given by:
\begin{eqnarray}
\langle e^{-\beta w} \rangle_{M} = \sum_k  \prod_{i=1}^N \langle e^{-\beta \dot{\lambda}(t_i) h \hX} \rangle_{k,M}\bra{\varphi_{k}} \rho_0 \ket{\varphi_k} \label{eq:expavgstep1},
\end{eqnarray}
where $\langle f(\hX) \rangle_{k,M} = \sum_{\alpha_j} f(x_{\alpha_j}) p(\alpha_j|\alpha_k)$. Applying the Jensen inequality with respect to the measurement distribution, $\langle e^{A}\rangle_{k,M}\geq e^{\langle A \rangle_{k,M}}$, to each term in the product on the right-hand side of (\ref{eq:expavgstep1}) and taking the limit $h \rightarrow 0$ we get
\begin{eqnarray*}
\langle e^{-\beta w} \rangle_M \geq \sum_{k} \exp \parlr{-\beta \parlr{\lambda(\tau)-\lambda(0)} \langle \hX \rangle_{k,M}} \bra{\varphi_{\alpha_k}} \rho_0 \ket{\varphi_{\alpha_k}}.
\end{eqnarray*}
Finally if we assume that we have a homogeneous distribution for $p(\alpha_j|\alpha_k)$, dependent only on the difference of the eigenvalues $\{x_{\alpha_{j}},x_{\alpha_k}\}$ and centered about $x_{\alpha_k}$ for each $k$, the mean $\langle X \rangle_{k,M}$ will equal $x_{\alpha_k}$ leading to
\begin{eqnarray*}
\langle e^{-\beta w} \rangle_M \geq \sum_{k} \exp \parlr{-\beta \parlr{\lambda(\tau)-\lambda(0)} x_{\alpha_k}} \bra{\varphi_{\alpha_k}} \rho_0 \ket{\varphi_{\alpha_k}}.
\end{eqnarray*}
Comparing the above equation to  (\ref{eq:expbetawpowinf}), we have the inequality
\begin{eqnarray}
\langle e^{-\beta w} \rangle_M \geq \langle e^{-\beta w} \rangle_p,\label{eq:inequalitygenmeas}
\end{eqnarray}
that is the exponential average of power-based work obtained from a generalised measurement of power is always greater than the one obtained from projective power measurements.
\section{Continuous measurement stochastic master equation}
\label{app:smederiv}
Let us consider the state of a system $\rho(t)$ that is normalized at time $t$ and is subject to the Gaussian measurement introduced in (\ref{eq:Gaussianmeas}) and with outcome $\alpha_t$ and a unitary evolution operator $U_t$ for a short interval $h$. We also chose the variance of the Gaussian measurement as $\sigma^2 = 1/(8\kappa h)$. The normalised state at $t+h$ is given by
\begin{eqnarray}
\rho(t+h) = \frac{M_{\alpha_t} U_t \rho(t) U_t^{\dagger} M_{\alpha_t}}{\Tr(M_{\alpha_t} U_t \rho(t) U_t^{\dagger} M_{\alpha_t})} \,\,. \label{eq:postmeas}
\end{eqnarray}
Examining the probability distribution for the measurement outcome, given by the denominator in (\ref{eq:postmeas}), as shown in\cite{Jacobs06} one can write (provided $h$ is chosen small enough so that the Gaussian is broader than the $\hX$ distribution of the state $\rho(t)$)
\begin{eqnarray}
\alpha_t = \langle \hX \rangle  + \frac{\Delta W}{\sqrt{8\kappa} h}, \label{eq:measout}
\end{eqnarray}
where $\Delta W$ is a zero mean Gaussian random variable with variance $h$ and $\langle \hX \rangle = \Tr (\rho(t) \hX)$. Taking into account the scaling $\Delta W^2 \sim h $ and (\ref{eq:measout}), we expand the operators in (\ref{eq:postmeas}) as
\begin{eqnarray}
U_t &=& I - i\frac{h}{\hbar} H_t + O(h^2),\label{eq:unitexp}\\
M_{\alpha_t} &\propto & I - \kappa(2 h-\Delta W^2) \parlr{\hX - \langle \hX\rangle}^2 + \sqrt{2\kappa}\Delta W \parlr{\hX - \langle \hX \rangle} \nonumber\\&+& O(h^{3/2}) \label{eq:measexp},
\end{eqnarray}
where we have denoted the instantaneous Hamiltonian as $H_t$. Substituting (\ref{eq:unitexp}) and (\ref{eq:measexp}) we get
\begin{eqnarray}
\rho(t+h) &=& \rho(t) - i\frac{h}{\hbar}[H_t,\rho(t)] + \sqrt{2\kappa} \Delta W \flrlr{\hX-\langle \hX \rangle,\rho(t)} \label{eq:rhoexp}
\\&-&\kappa (2 h-\Delta W^2)\flrlr{\hX^2,\rho(t)} + 2\kappa\Delta W^2 \hX \rho(t) \hX \nonumber\\
&-& 4 \kappa (\Delta W^2-h) \Tr \parlr{\hX-\langle \hX \rangle}^2 \rho(t) + O(h^{3/2}) \nonumber.
\end{eqnarray}
Taking the infinitesimal limit $h \rightarrow dt$ and $\Delta W \rightarrow dW = \xi dt$, the Wiener differential in (\ref{eq:rhoexp}) and observing that $dW^2 = dt$, we immediately obtain (\ref{eq:contsme}).

\end{document}